\documentclass[pdflatex,sn-vancouver-num]{sn-jnl}

\usepackage{graphicx}%
\usepackage{multirow}%
\usepackage{amsmath,amssymb,amsfonts}%
\usepackage{mathrsfs}%
\usepackage{xcolor}%
\usepackage{textcomp}%
\usepackage{manyfoot}%
\usepackage{booktabs}%
\usepackage{algorithm}%
\usepackage{algorithmicx}%
\usepackage{algpseudocode}%
\usepackage{listings}%
\usepackage{array}%
\usepackage{url}

\newcommand{\orcidtext}[1]{}
\hypersetup{
  hypertexnames=false,
  pdftitle={Sharp Target-Domain Certificates for Quantum-Kernel Advantage under Distribution Shift},
  pdfauthor={Roberto Fernandez-Barrios, Iker Pastor-Lopez, Asier Gonzalez-Santocildes, Pablo Garcia Bringas},
  pdfsubject={Research article submitted to EPJ Quantum Technology},
  pdfkeywords={quantum machine learning, quantum kernels, quantum advantage, partial identification, active testing, distribution shift},
  hidelinks
}

\theoremstyle{thmstyleone}%
\newtheorem{theorem}{Theorem}
\theoremstyle{thmstyletwo}%

\theoremstyle{thmstylethree}%

\begin{document}

\title[Certificates for Quantum-Kernel Advantage]{Sharp Target-Domain Certificates for Quantum-Kernel Advantage under Distribution Shift}


\author*[1]{\fnm{Roberto} \sur{Fernández-Barrios}\orcidtext{0009-0003-5312-2634}}\email{roberto.fernandez.b@deusto.es}

\author[1]{\fnm{Iker} \sur{Pastor-López}\orcidtext{0000-0002-3068-6248}}\email{iker.pastor@deusto.es}

\author[1]{\fnm{Asier} \sur{González-Santocildes}\orcidtext{0009-0002-8888-8560}}\email{gonzalez.asier@deusto.es}

\author[1]{\fnm{Pablo} \sur{García Bringas}\orcidtext{0000-0003-3594-9534}}\email{pablo.garcia.bringas@deusto.es}

\affil*[1]{\orgdiv{Faculty of Engineering}, \orgname{University of Deusto}, \orgaddress{\street{Avda. de las Universidades, 24}, \city{Bilbao}, \postcode{48007}, \country{Spain}}}


\abstract{Quantum predictive advantage under shift usually requires known target labels. We derive the assumption-free sharp identified interval for the finite-batch advantage of a fixed candidate over the best member of a prespecified fixed classical-kernel family under any bounded loss and unrestricted completions of the unaudited labels; zero--one accuracy has exact closed-form partial-label updates. Across eight security shifts, zero-label upper endpoints against 115 classical kernels span 0.002--0.088. Retrospective auditing reduces every endpoint to 0.010 with 0--33 of 500 labels, including entangling-ZZ models. Prospective corroboration on two eligible tasks met predefined criteria: four task--classifier medians require 0--3 labels (overall median 0; maximum 76), and all 20 realized effects are negative; a third task fails its feature gate. Protocol controls reverse an apparent advantage, while finite-shot noise can increase predictive distinctness without useful advantage. The framework separates protocol validity, target predictive indispensability, and quantum relevance.}




\maketitle

\section{Introduction}
\label{sec:introduction}

Quantum kernel methods embed classical data into quantum states and use state overlaps as similarities for otherwise classical learners \cite{havlicek2019quantumenhanced,schuld2019featurehilbert,schuld2021kernelmethods}. This separation makes them a clean setting for studying quantum inductive bias, but a predictive-advantage claim must pass three distinct tests. The comparison protocol must be valid; the target predictions must resist reproduction by an appropriate classical reference family; and the surviving behaviour must depend on a non-trivial quantum construction whose resources are counted. A model can pass one test and fail another.

This article therefore studies a deliberately narrow object: for a fixed quantum candidate and a prespecified classical-kernel reference family, the set of predictive advantages on the actual deployment batch that remain compatible with the target labels not yet inspected. We make four contributions. First, for any additive bounded loss on a finite label space, we derive the interval hull of the partial-label identified set for advantage over the best member of a prespecified classical-kernel reference family and show that no assumption-free rule using the same information can return a uniformly smaller valid interval. Zero--one accuracy has an unusually simple closed form and is pathwise exact after any realized audit subset, including subsets selected adaptively. Second, nested classical reference families and target-label audits define a reference-breadth--target-supervision evidence frontier. Third, we give an exact prevalence-conditional balanced-accuracy formulation and an explicit positive-semidefinite (PSD) construction showing why train-only geometry cannot determine target predictions. Fourth, we embed these objects in a controlled quantum-kernel benchmark and prospectively specify a Gate~2 replication before opening target labels; entanglement strata and repeated finite-shot perturbations separately test when predictive distinctness is quantum-relevant.

\begin{figure*}[t]
\centering
\includegraphics[width=\textwidth]{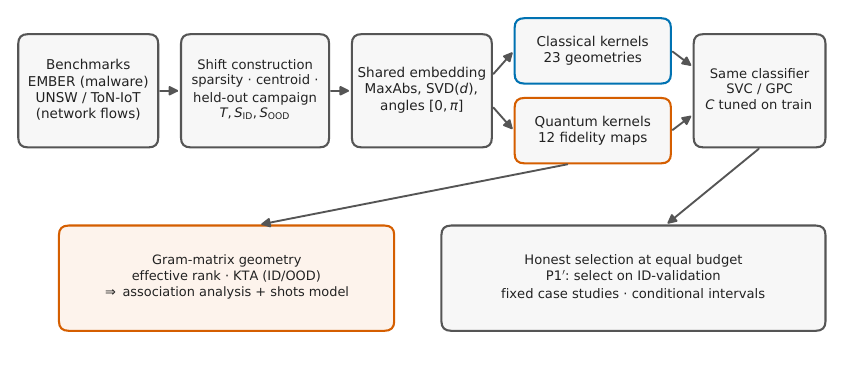}
\caption{The controlled kernel-swap protocol. Benchmark pools feed reproducible shift constructions; all kernels operate on a shared train-only embedding, receive per-configuration regularization tuned on the training split, and are consumed by the same precomputed-kernel classifiers. Family-level conclusions are drawn by target-label-free, equal-budget selection on an in-distribution validation split (P1$'$). Each scenario-group interval is a 95\% conditional cluster-$t$ interval over five q-split seed clusters ($n=5$), conditional on the fixed benchmark pools and design. Gram matrices feed the associational geometry audit and repeated finite-shot sensitivity.}
\label{fig:protocol}
\end{figure*}

The benchmark uses four security scenarios from EMBER, UNSW-NB15, and ToN-IoT, producing eight fixed constructed or held-out-campaign shifts. Support-vector classifiers (SVC) and Laplace Gaussian-process classifiers (GPC) consume identical precomputed Gram matrices. The quantum family contains entangling-ZZ and separable-product fidelity maps at 4--12 qubits; the classical family contains 115 linear, radial-basis-function (RBF), polynomial, Laplacian, and Mat\'ern candidates. Models are tuned on training data and selected on a disjoint in-distribution (ID) validation split (Figure~\ref{fig:protocol}). Equal 60-candidate pools support fair performance comparison, while the complete prespecified classical-kernel reference family supports the separate Gate~2 indispensability audit. One prospectively specified three-task TableShift study corroborates protocol sensitivity; a second tests Gate~2 prospectively on previously unused task identifiers.

Against the full prespecified classical-kernel reference family, the zero-label upper endpoint of accuracy advantage ranges from 0.002 to 0.054 for SVC and 0.006 to 0.088 for GPC. In a retrospective fixed-batch analysis, an adaptive bottleneck-coverage audit reduces it to at most 0.010 using 0--33 of 500 labels (median 13 over 16 fixed case--classifier cells). Dynamic random-disagreement and non-adaptive initial-coverage controls both have median 12.5, so the large gain over prediction-independent ordering is attributable to excluding non-actionable queries rather than to the proposed coverage heuristic. The result persists for selected entangling-ZZ models. Prospective corroboration on two technically eligible tasks met the predefined criteria: four task--classifier medians were 0--3 labels, the overall seed-level median was 0, and the maximum was 76. All 20 prospective realized accuracy effects were negative. The third task failed its prespecified minimum-feature gate before model execution. The certificate remains conditional on the prespecified classical-kernel reference family and target batch, not evidence of classical simulability or complexity-theoretic separation. It converts the earlier benchmark conclusion---that controls remove an apparent gain---into a quantitative question: how much advantage remains possible, and how much classical search and target supervision suffice to falsify a material claim?

\subsection{Relation to prior work}
\label{sec:intro_prior}

Recent benchmarks emphasize the first test. Broad quantum machine-learning (QML) comparisons find competitive classical baselines once pipelines are aligned \cite{bowles2024betterthanclassical}; Slattery et al. show numerically that general-purpose hyperparameter tuning can make fidelity kernels well approximated by classical kernels on classical data \cite{slattery2023numerical}; and quantum-kernel results depend strongly on feature maps and hyperparameters \cite{schnabel2025scrutiny}; and a recent intrusion-detection audit with tuned classical baselines, calibration-aware metrics, and a simulated noise sweep finds that tuned classical models match or exceed hybrid and quantum-kernel models on aggregate detection, with only two narrow exceptions surviving multiplicity correction \cite{gillani2026howquantum}. Those studies concern benchmark performance or global kernel approximation. Our contribution instead gives an exact finite-target interval conditional on fixed predictions and a prespecified classical-kernel reference family, including under partial target-label auditing. A concurrent finance preprint likewise shows an apparent quantum-kernel advantage vanishing under point-in-time, kernel-swap, and budget controls \cite{shen2026vanishing}. Under distribution shift the selection rule becomes part of the estimand: same-test selection is an oracle, train-only regularization can change rankings, and a larger candidate family has more opportunities to win \cite{gulrajani2021domainbed,koh2021wilds,yao2022wildtime,zhou2023dgsurvey}. Fair equal-budget benchmarking answers who wins under symmetric search, but not whether any classical witness in a declared family can reproduce the selected quantum model on the deployment inputs.

Geometric and surrogate analyses approach that second question from complementary directions. Huang et al. use a label-independent geometric difference to screen whether a quantum kernel can have a prediction advantage \cite{huang2021powerofdata}; classical-surrogate constructions reproduce input--output relations of trained quantum models \cite{schreiber2023surrogates}; a closed-form classical surrogate has been derived for the ZZ feature map in the small-bandwidth regime \cite{tekeli2026zzsurrogate}; and recent theory studies when quantum learners depart from minimum-norm classical solutions \cite{thabet2026disagree}. Those objects characterize geometric potential, global approximation, or learning-rule separation. Our target-batch question is different. A train-only Gram matrix does not determine how a kernel transports predictions to a shifted target because deployment also depends on the target--train block, while different geometries can yield identical decisions on the observed target inputs. We therefore audit predictive indispensability directly on target covariates while keeping their labels protected.

That protection creates an identification problem in the tradition of partial identification: observed information and maintained restrictions can determine a set of compatible values without point-identifying the target \cite{manski2003partialidentification,molinari2020microeconometrics}. Here the observed object is unusually concrete---one finite batch and fixed prediction matrices---and the maintained label restriction is deliberately empty. The estimand is the relative functional ``fixed candidate versus best member of a prespecified fixed family.'' We derive its sharp interval over unrestricted completions of the unaudited labels, rather than estimate a population parameter or add assumptions until point identification is recovered. Weak-supervision work also uses partial identification, but combines fitted label-model marginals with Fr\'echet restrictions for individual metrics \cite{polo2024partialidentification}; those marginals are neither available nor required here. Classifier evaluation with missing labels has also been addressed by multiple imputation, which yields point estimates and a predictive distribution for metrics such as precision, recall, and ROC-AUC \cite{dervovic2025model}; our object is instead the exact finite-batch identified interval over all completions of the unobserved labels for a relative best-of-family advantage.

Label-efficient evaluation pursues a different objective. Active evaluation has used adaptive stratified sampling \cite{katariya2012activeevaluation}, importance-weighted Active Testing \cite{kossen2021activetesting}, learned loss surrogates \cite{kossen2022activesurrogate}, and prediction-powered control variates \cite{ashouritaklimi2026predictionpowered} to estimate risk accurately with fewer labels. Limited-label model selection instead chooses which pretrained classifier to deploy \cite{okanovic2025modelselector}, while semi-supervised multi-model evaluation borrows information across classifier scores and unlabeled data under a mixture model \cite{shanmugam2025evaluating}. Under distribution shift, average thresholded confidence (ATC) predicts target accuracy from confidence \cite{garg2022leveraging}; disagreement-discrepancy methods provide assumption-dependent error bounds \cite{rosenfeld2023disagreement,mishra2025odd}; and label-complexity results show that, under joint covariate--label shift, no label-free method is simultaneously per-class valid and efficient uniformly over the target laws consistent with the source data and target covariates \cite{han2026labelcomplexity}. Target risk under simultaneous covariate shift and selectively observed outcomes has also been identified under explicit conditions and estimated by double machine learning \cite{ulichney2026beyond}. Our certificate estimates neither labels nor risk: it retains every completion and reports the exact range of a fixed relative functional. Co-validation and discordant-pair evaluation remain complementary uses of disagreement \cite{madani2004covalidation,musgrove2023discordant}.

The distinction from minimax transductive learning is equally important. Balsubramani and Freund formulate ensemble prediction on an unlabeled test set as a game, using prior classifier--label correlation constraints to choose a new predictor that minimizes worst-case error \cite{balsubramani2015optimally}. We do not optimize or combine predictions and impose no such label constraints. The candidate and reference predictions remain fixed; the adversarial completion is used only to identify the sharp range of their realized finite-batch advantage. Thus transductive minimax aggregation is a decision rule under supplied performance information, whereas our object is an assumption-free identification certificate conditional on a declared prediction matrix. The distinction concerns the target, admissible information, and action, not only the computational form.

A companion study by the present authors addresses a different question in a different domain: the conditional validity of quantum event classifiers under collider systematics and finite-shot estimation uncertainty, using an information-conditional, fail-closed auditing framework with anytime-valid per-claim error control \cite{fernandezbarrios2026conditional}. It shares neither datasets, estimands, nor guarantees with the finite-batch certificate developed here.

\section{Results}
\label{sec:results}

\subsection{Overview}

The results are organized as three gates. Gate~1 asks whether the comparison protocol itself is valid: regularization is fitted on training data, deployment candidates are selected without out-of-distribution (OOD) labels, and family search budgets are distinguished from unrestricted adversarial reference breadth. Gate~2 asks whether the selected quantum model is predictively indispensable relative to the prespecified classical-kernel reference family on the observed target inputs. This is the new target-batch certificate and evidence frontier. Gate~3 asks whether any surviving behaviour is quantum-relevant, separating entangling-ZZ maps from product maps and accounting for finite-shot estimation. The eight security groups are fixed cases; conditional intervals over q-split clusters describe pipeline variation given the benchmark pools and design, not uncertainty over a population of datasets. Extended results, complete sensitivity tables, and diagnostics are provided as Supplementary Information (Additional file~1).

\subsection{Sharp target-batch identification of predictive advantage}
\label{sec:res_certificate}

Consider a finite target batch, a fixed quantum prediction, and $M$ fixed classical predictions. For a finite label space $\mathcal Y$ and an additive bounded loss, let $\ell_i^Q(y)$ and $\ell_{ij}^C(y)$ be their losses on item $i$ if its label were $y$, and define the classical-minus-quantum contrast $a_{ij}(y)=\ell_{ij}^C(y)-\ell_i^Q(y)$. Positive advantage favours the quantum model:
\[
\Delta_\ell(y)=\min_{j\leq M}\frac1n\sum_{i=1}^n a_{ij}(y_i).
\]
If labels have been audited on $L$ and $S_j(L)=\sum_{i\in L}a_{ij}(y_i)$, every assignment on $L^c$ remains admissible.

\begin{theorem}[Sharp bounded-loss interval hull: sharpness and minimality]
\label{thm:bounded_loss}
The smallest interval containing $\Delta_\ell(y)$ for every completion of the unaudited labels is $[L_\ell(L),U_\ell(L)]$, where
\[
L_\ell(L)=\frac1n\min_j\left\{S_j(L)+
\sum_{i\notin L}\min_{y\in\mathcal Y}a_{ij}(y)\right\},
\]
\[
U_\ell(L)=\frac1n\max_{y_{L^c}\in\mathcal Y^{|L^c|}}
\min_j\left\{S_j(L)+\sum_{i\notin L}a_{ij}(y_i)\right\}.
\]
Both endpoints are attained. Consequently, any assumption-free interval computed from the same fixed losses and audited labels and valid for every admissible completion must contain both endpoints.
\end{theorem}

The lower endpoint separates over labels after choosing a witness. The upper endpoint is an exact finite max--min problem, expressible as a mixed-integer linear program (MILP) with one-hot label variables. Auditing any additional label removes completions, so the lower endpoint cannot decrease and the upper endpoint cannot increase. This general statement is not a claim that loss-based partial identification or active evaluation is new; the contribution is the exact relative-family estimand and its sharp, minimal finite-batch certificate. Zero--one classification has further structure: assigning an unaudited label to the quantum hard prediction simultaneously maximizes its contrast with every disagreeing classical witness. This yields the following closed forms used in all experiments.

Fix the hard target predictions of a selected quantum classifier $h_Q$ and a prespecified classical family $\mathcal C_B=\{h_{C_1},\ldots,h_{C_M}\}$ on $n$ target inputs. For an otherwise unrestricted target labelling $y$, define
\[
\Delta_B(y)=\operatorname{Acc}(h_Q;y)-
             \max_{j\leq M}\operatorname{Acc}(h_{C_j};y),
\qquad
d_j=\frac{1}{n}\sum_{i=1}^{n}
\mathbf 1\{h_Q(x_i)\neq h_{C_j}(x_i)\}.
\]

\begin{theorem}[Sharp zero-label region]
\label{thm:zero_label}
The smallest interval containing $\Delta_B(y)$ for every possible target labelling is
\[
\mathcal I_B^{(0)}
=\left[-\max_{j\leq M}d_j,\ \min_{j\leq M}d_j\right].
\]
Both endpoints are attainable, including for multiclass hard predictions.
\end{theorem}

The upper endpoint is not a probabilistic bound or an estimate: it is the exact largest advantage compatible with the observed prediction matrix when every target label remains unknown. It is attained by setting $y=h_Q$. The lower endpoint is attained by assigning the predictions of the classical model farthest from $h_Q$. Hence no smaller label-free interval is valid without extra assumptions. In particular, if $\min_j d_j<\tau$, any accuracy advantage of at least $\tau$ over this classical family is falsified on the batch, even before a target label is revealed. Whenever $\min_jd_j>0$, the same target covariates and prediction matrix remain compatible with both signs of advantage; disagreement alone cannot identify which family wins.

\begin{figure*}[t]
\centering
\includegraphics[width=\textwidth]{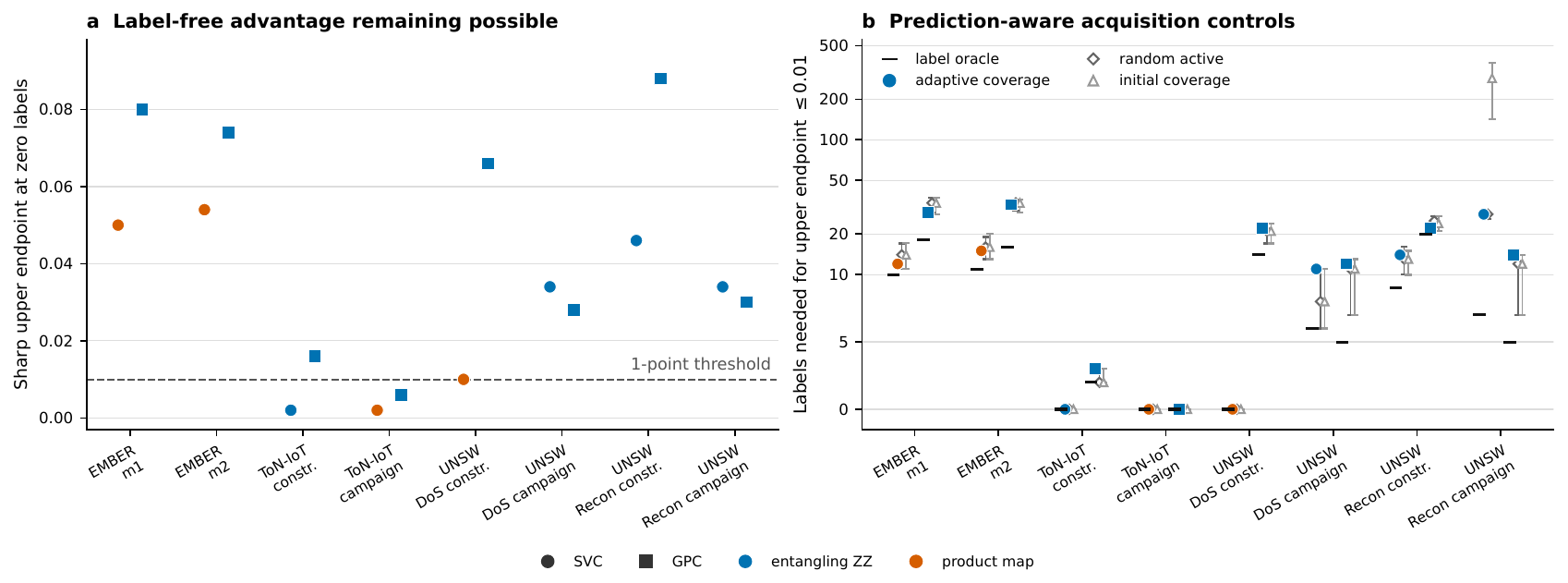}
\caption{Sharp target-batch certificates against the full 115-candidate classical family. \textbf{a}, Zero-label upper endpoints for the SVC and GPC quantum models fixed by P1$'$. Colour separates selected entangling-ZZ and product-map models; the dashed line marks 0.010 accuracy. \textbf{b}, Labels needed to reduce the endpoint to at most 0.010 (symmetric-log scale). Horizontal ticks are the exact retrospective label oracle; filled markers are the fixed adaptive coverage rule; open diamonds and triangles are medians for dynamic random-active and non-adaptive initial-coverage controls, with bars spanning 2.5--97.5\% across 200 SHA-256 tie orders. All certificates use the canonical q1000 realization (master, q-split, and model seeds 42) and a 500-point target batch. Comparator ranges are tie-order sensitivities, not confidence intervals.}
\label{fig:v9_certificate}
\end{figure*}

The zero-label certificate is already restrictive in several cases (Figure~\ref{fig:v9_certificate}a). Against the complete 115-candidate classical family, the upper endpoint ranges from $0.002$ to $0.054$ for SVC and from $0.006$ to $0.088$ for GPC (Table~\ref{tab:v9_certificate}). A small endpoint says that at least one classical witness nearly reproduces the selected quantum decisions; it does not say that the quantum kernel is efficiently simulable or globally reproducible.

\begin{theorem}[Sharp arbitrary-partial-label region]
\label{thm:partial_label}
Let $L$ be any audited subset. For classical witness $j$, let
\[
S_j(L)=\sum_{i\in L}
\left[\mathbf 1\{h_Q(x_i)=y_i\}-
      \mathbf 1\{h_{C_j}(x_i)=y_i\}\right]
\]
and let $R_j(L)$ count unaudited points on which $h_Q$ and $h_{C_j}$ disagree. Then the sharp identified interval over all completions of the unaudited labels is
\[
\mathcal I_B(L)=
\left[
\min_j\frac{S_j(L)-R_j(L)}{n},\
\min_j\frac{S_j(L)+R_j(L)}{n}
\right].
\]
The upper endpoint is non-increasing as labels are audited and collapses to the realized family advantage when $L$ contains the full batch.
\end{theorem}

We use this result as an active falsification audit, not as a new model-selection rule. At each step the models remain fixed; one tested rule queries the point covering the largest number of witnesses currently tying at the smallest reducible upper bound (Methods). In this retrospective fixed-batch analysis, the rule reduces the $0.010$ endpoint using 0--33 of 500 labels, with median 13 across the 16 case--classifier cells (Figure~\ref{fig:v9_certificate}b). Stronger controls remove any claim that the coverage heuristic itself is unusually efficient: dynamic random sampling restricted to current actionable disagreements and a fixed initial-coverage ordering both have median 12.5 labels. Against random-active sampling, the adaptive rule wins, ties, and loses in 5, 5, and 6 cells; against initial coverage the counts are 6, 4, and 6. An exact ex-post label oracle needs median 6.5 labels, and the adaptive rule's median excess is 4.5. The main acquisition result is therefore that prediction-aware restriction makes a small audit possible, not that one heuristic dominates. Prediction-independent ordering remains a weak diagnostic (median 245.5) and is reported only in Supplementary Information.

\begin{table*}[t]
\centering
\caption{Sharp accuracy certificates for the eight fixed target batches against the full 115-candidate classical family. $U_0$ is the zero-label upper endpoint; $n_{.01}$ is the fixed adaptive-coverage count for $U\leq0.010$; $\Delta_{\rm full}$ is the point-identified quantum advantage after all 500 labels are unlocked. The canonical run, quantum model, and every classical candidate were fixed before this audit; target labels are never used to retune or reselect them.}
\label{tab:v9_certificate}
\scriptsize
\setlength{\tabcolsep}{4pt}
\begin{tabular}{@{}lrrr@{\qquad}rrr@{}}
\toprule
& \multicolumn{3}{c}{SVC} & \multicolumn{3}{c}{GPC}\\
\cmidrule(lr){2-4}\cmidrule(lr){5-7}
Target case & $U_0$ & $n_{.01}$ & $\Delta_{\rm full}$ &
$U_0$ & $n_{.01}$ & $\Delta_{\rm full}$\\
\midrule
EMBER $m1$                 & .050 & 12 & $-.024$ & .080 & 29 & $-.030$\\
EMBER $m2$                 & .054 & 15 & $-.036$ & .074 & 33 & $-.016$\\
ToN-IoT constructed        & .002 &  0 & $-.004$ & .016 &  3 & $-.014$\\
ToN-IoT campaign           & .002 &  0 & $-.018$ & .006 &  0 & $-.012$\\
UNSW DoS constructed       & .010 &  0 & $-.016$ & .066 & 22 & $-.052$\\
UNSW DoS campaign          & .034 & 11 & $-.026$ & .028 & 12 & $-.032$\\
UNSW Recon constructed     & .046 & 14 & $-.032$ & .088 & 22 & $-.066$\\
UNSW Recon campaign        & .034 & 28 & $-.026$ & .030 & 14 & $+.002$\\
\bottomrule
\end{tabular}
\end{table*}

\subsection{A reference-breadth--target-supervision evidence frontier}
\label{sec:res_frontier}

Nested classical reference families and partial labels expose two different costs of falsification. Increasing the classical family can only decrease the certificate's upper endpoint; auditing informative target labels can only decrease it further. We therefore report the surface $(B,|L|)\mapsto U_B(L)$ and, for a materiality threshold $\tau$, the smallest reference breadth and target supervision required to establish $U_B(L)\leq\tau$. Here $B$ indexes a prespecified \emph{reference breadth}, not runtime or computational complexity.

The frozen block-ordering analysis already provides three nested breadths: 30, 60, and 115 candidates. For each cell we take the median zero-label endpoint over 5,000 prespecified whole-block orderings, then summarize the 16 case--classifier cells. The cross-cell median contracts from 0.042 at $B=30$ to 0.034 at $B=60$ and remains 0.034 at $B=115$; the number of cells already at or below 0.010 changes from 3 to 4 to 4 (Supplementary Table~22). Thus most aggregate contraction occurs by 60 candidates, but the full family remains necessary for the deliberately adversarial question and for individual cases. These ordering distributions quantify dependence on reference breadth; they are not confidence intervals or computational-cost measurements.

The prediction-aware evidence curves contract rapidly because only disagreements capable of lowering a current witness endpoint need adjudication (Supplementary Figure~1). The adaptive rule is not consistently better than random-active sampling, although recomputing the active set avoids the 286.5-label median outlier produced by a frozen initial-coverage ordering. The result persists when the selected quantum model uses an entangling-ZZ map: across the 12 entangling case--classifier cells, the adaptive rule needs 0--33 labels (median 14) for the 0.010 endpoint, while random-active cell medians are 0--34 (median 12.5). Thus the finding is not an artefact of including classically factorizable product maps.

Reference breadth matters separately from fair benchmarking. For the ToN-IoT constructed shift under GPC, the customary 30-candidate linear/RBF family yields a realized quantum accuracy advantage of $+0.028$, so no number of labels can falsify a one-point claim against that restricted reference. Against the full 115-candidate family the realized advantage is $-0.014$, the zero-label upper endpoint is $0.016$, and three targeted labels reduce it below $0.010$. Equal 60-versus-60 budgets remain the appropriate performance comparison; the full family asks the different, deliberately adversarial question of predictive indispensability.

\subsection{Prospective Gate-2 corroboration against the prespecified classical-kernel reference family}
\label{sec:res_gate2_prospective}

The security acquisition curves above are retrospective. We therefore specified a separate Gate-2 replication against the prespecified classical-kernel reference family before acquiring its three TableShift tasks, computing a model prediction, or opening a target label. The task identifiers were BRFSS Diabetes, ACS Food Stamps, and NHANES Lead. The immutable specification fixed five label-blind sampling seeds, the q1000 split sizes, P1$'$ selection, both classifiers, the 60-quantum and 115-classical pools, physical prediction--label separation, three thresholds, 200 SHA-256 orders for each strong control, and numerical outcome criteria. The aggregate prediction lock was hashed before the one-time label audit (Methods).

All five NHANES seeds produced only 11 non-constant post-encoding features, below the frozen 12-dimensional grid requirement. The task therefore failed a prespecified minimum-feature gate before any model candidate was executed; it was not replaced or repaired by reducing the grid. The replication is consequently technically limited to ten task--seed units, or 20 fixed task--seed--classifier cells, rather than evidence from three complete tasks.

\begin{figure*}[t]
\centering
\includegraphics[width=\textwidth]{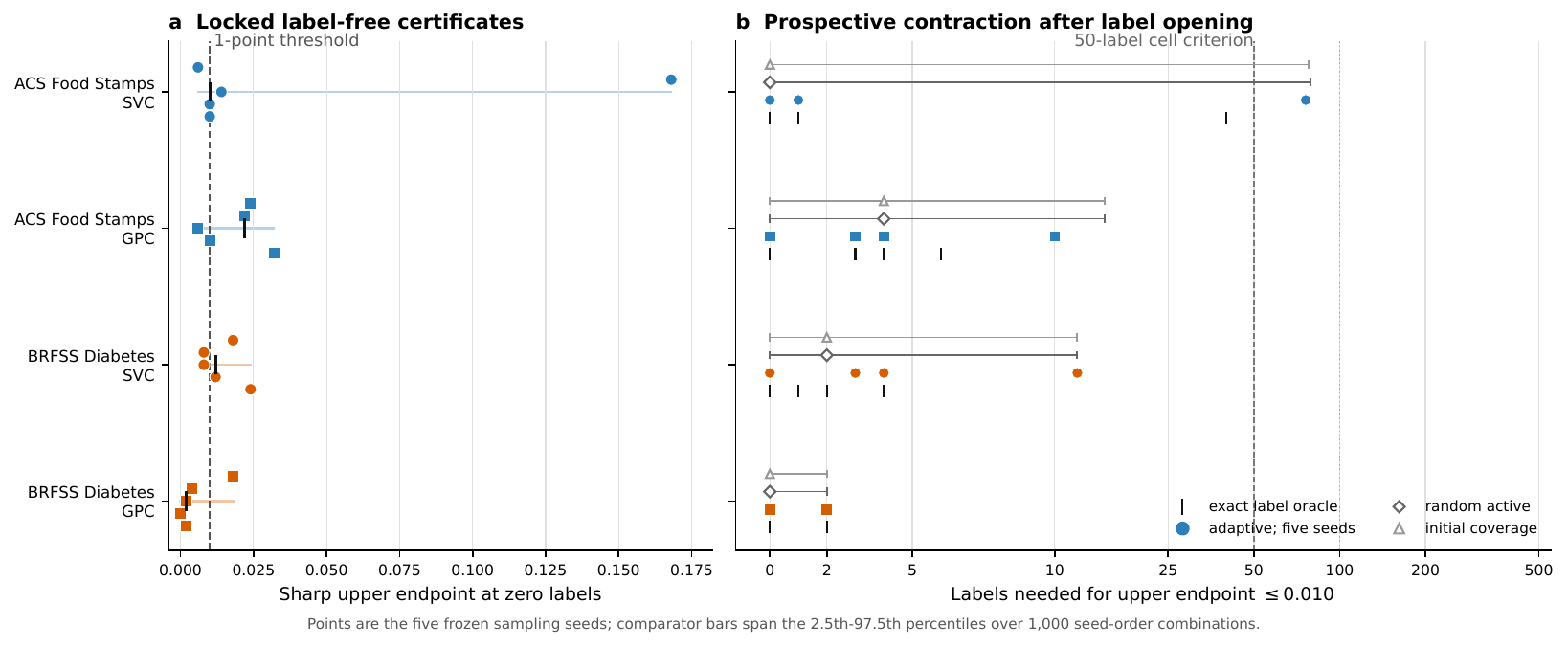}
\caption{Prospectively specified Gate-2 replication against the full 115-candidate prespecified classical-kernel reference family. \textbf{a}, Zero-label upper endpoints for five deterministic sampling seeds in each technically eligible task--classifier cell; black ticks mark medians and the dashed line marks 0.010 accuracy. \textbf{b}, Labels required to reduce the endpoint to at most 0.010. Coloured points are the five fixed adaptive counts, black ticks are exact retrospective label-oracle counts, and open markers with bars are medians and 2.5--97.5\% quantiles for dynamic random-active and non-adaptive initial-coverage policies across 1,000 seed--order combinations (five seeds times 200 SHA-256 orders). The 50-label line is the predefined task--classifier median criterion; the dotted 100-label line is the predefined seed-level ceiling. Comparator ranges are algorithmic sensitivities, not confidence intervals. NHANES Lead failed the prespecified 12-feature gate before model execution and is absent.}
\label{fig:v10_gate2}
\end{figure*}

The two technically eligible tasks met the predefined numerical criteria (Figure~\ref{fig:v10_gate2}). Against the full prespecified classical-kernel reference family, zero-label endpoints range from 0 to 0.168 (median 0.010), and 11/20 cells already satisfy $U\leq0.010$ without target labels. The fixed adaptive audit reduces every remaining endpoint to that threshold with 0--76 labels (overall median 0). The four task--classifier medians are 0 for ACS SVC, 3 for ACS GPC, 3 for BRFSS SVC, and 0 for BRFSS GPC; all are below the 50-label criterion, and the worst seed remains below the predefined 100-label ceiling. Every realized accuracy effect is negative, ranging from $-0.156$ to $-0.002$.

The prospective controls reproduce the earlier negative algorithmic conclusion. Against both dynamic random-active and non-adaptive initial-coverage medians, adaptive coverage wins, ties, and loses in 3, 15, and 2 of the 20 cells, respectively; the median paired difference is zero. Prediction-independent ordering is far less efficient in the non-trivial cells. Thus the transferred result concerns the sharp certificate and restriction to actionable disagreements, not superiority of the coverage heuristic. All prospectively selected quantum winners are product maps, so this replication does not independently validate the entangling-ZZ stratum; that stratum remains supported only by the retrospective security cases.

\subsection{The apparent advantage under oracle selection}
\label{sec:res_baselines}

We begin with a deliberately optimistic reference comparison: fidelity-based quantum kernels against linear and RBF baselines, with each family's representative chosen by Best-by-OOD selection---the configuration with the highest balanced accuracy on the OOD test split itself---and the classical kernels left at the default regularization $C=1$. Related intrusion-detection studies motivate the small-reference context \cite{abreu2024qmlids,abreu2025quantumnetsec,bellante2025evaluating}, but we do not claim that this exact protocol dominates QML practice. Under these conventions, on EMBER, the quantum family shows an OOD advantage of $+0.037$ under SVC and $+0.028$ under GPC against linear+RBF, shrinking to $+0.004$/$+0.002$ once the classical family is enlarged with heavier-tailed kernels. Supplementary Table~12 gives the complete oracle results.

Two features favour this apparent quantum result. Best-by-OOD is an \emph{oracle}: it uses the labels it then reports. Fixing $C=1$ also leaves classical candidates at a potentially unsuitable regularization point. These values are therefore descriptive upper bounds, not deployable effects.

\subsection{The controlled comparison does not preserve the apparent advantage}
\label{sec:res_honest}

A deployable comparison must tune each configuration's regularization without looking at the test data, select the deployed configuration without OOD labels, and compare families at the same candidate budget. We implement all three (Section~\ref{sec:methods}). For every kernel candidate, the SVC regularization $C$ is chosen by five-fold cross-validation on the training Gram matrix alone. The deployed configuration is then the one with the best balanced accuracy on an in-distribution \emph{validation} split, held out from the ID test split (protocol P1$'$); no OOD label is ever consulted in selection. The primary comparison uses 60 candidates per extended family in every scenario-group and classifier, with the full 115-candidate classical pool retained as a sensitivity. We treat the eight scenario-groups as fixed case studies and report, per group, the quantum-minus-classical OOD delta with a conditional interval over the five split realizations, and no population $p$-value (Section~\ref{sec:methods}).

\begin{figure*}[t]
\centering
\includegraphics[width=\textwidth]{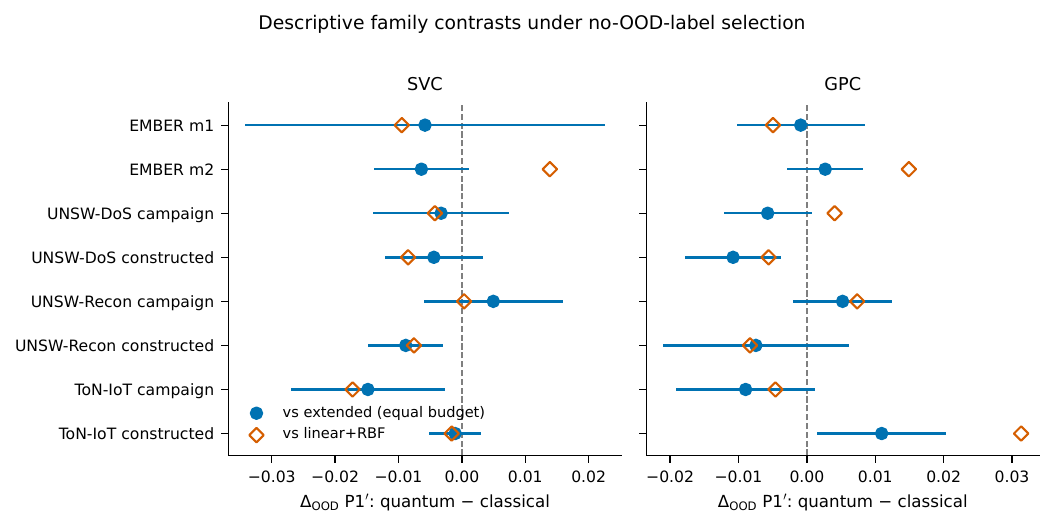}
\caption{Descriptive target-label-free family contrasts in eight fixed scenario-groups. P1$'$ OOD differences (quantum minus classical) are shown for SVC (left) and GPC (right). Filled circles compare against the equal-budget extended classical family; error bars are pointwise 95\% conditional cluster-$t$ intervals over five q-split clusters ($n=5$), conditional on the fixed benchmark pools and design. Open diamonds compare against the customary linear+RBF reference. The intervals are not simultaneous and do not imply population inference or equivalence.}
\label{fig:honest}
\end{figure*}

Table~\ref{tab:headline} and Figure~\ref{fig:honest} show no consistent evidence of a quantum advantage under the controlled endpoint. Against the equal-budget extended classical family, the source-dataset-equal difference is $-0.00565$ for SVC (leave-one-source-dataset-out range $[-0.00702,-0.00450]$) and $-0.00094$ for GPC ($[-0.00190,+0.00094]$). Against the customary linear+RBF reference, the corresponding differences are $-0.00407$ and $+0.00591$, with heterogeneous group signs. UNSW-NB15 reconnaissance under held-out-campaign shift has a small positive mean ($+0.005$ for both classifiers), with pointwise conditional intervals spanning zero. The full-pool extended-family sensitivity is similar in direction and scale ($-0.00378$ for SVC and $-0.00293$ for GPC).

\begin{table*}[t]
\centering
\caption{Target-label-free selection results for eight fixed scenario-groups. Values are quantum-minus-classical OOD balanced-accuracy differences under P1$'$ (select on ID-validation, tune per-configuration $C$ on train, evaluate on OOD test). The frozen extended-family analysis uses 60 candidates per family in every group--classifier cell. Each scenario interval is a 95\% conditional cluster-$t$ interval over five q-split seed clusters ($n=5$), conditional on the benchmark pools and design; the fifteen pipeline realizations per setting are nested within these clusters. An interval containing zero is not an equivalence result. Source-dataset-equal rows average groups within three source datasets and report leave-one-source-dataset-out (LODO) ranges, not confidence intervals. The customary linear+RBF comparison is contextual rather than equal-budget.}
\label{tab:headline}
\scriptsize
\setlength{\tabcolsep}{3pt}
\begin{tabular}{@{}llccc@{}}
\toprule
Scenario & Clf. & vs lin./RBF & vs ext. (equal $B$) & interval / LODO range \\
\midrule
\multirow{2}{*}{EMBER $m1$} & SVC & $-0.0095$ & $-0.0058$ & $[-0.0340, +0.0223]$ \\
 & GPC & $-0.0050$ & $-0.0009$ & $[-0.0101, +0.0083]$ \\
\addlinespace
\multirow{2}{*}{EMBER $m2$} & SVC & $+0.0139$ & $-0.0064$ & $[-0.0137, +0.0009]$ \\
 & GPC & $+0.0149$ & $+0.0027$ & $[-0.0027, +0.0081]$ \\
\addlinespace
\multirow{2}{*}{UNSW-DoS (campaign)} & SVC & $-0.0043$ & $-0.0033$ & $[-0.0139, +0.0073]$ \\
 & GPC & $+0.0041$ & $-0.0057$ & $[-0.0120, +0.0005]$ \\
\addlinespace
\multirow{2}{*}{UNSW-DoS (constructed)} & SVC & $-0.0085$ & $-0.0044$ & $[-0.0119, +0.0031]$ \\
 & GPC & $-0.0056$ & $-0.0108$ & $[-0.0177, -0.0039]$ \\
\addlinespace
\multirow{2}{*}{UNSW-Recon (campaign)} & SVC & $+0.0004$ & $+0.0050$ & $[-0.0058, +0.0157]$ \\
 & GPC & $+0.0074$ & $+0.0052$ & $[-0.0019, +0.0123]$ \\
\addlinespace
\multirow{2}{*}{UNSW-Recon (constructed)} & SVC & $-0.0076$ & $-0.0089$ & $[-0.0146, -0.0031]$ \\
 & GPC & $-0.0083$ & $-0.0075$ & $[-0.0208, +0.0059]$ \\
\addlinespace
\multirow{2}{*}{ToN-IoT (campaign)} & SVC & $-0.0172$ & $-0.0148$ & $[-0.0268, -0.0029]$ \\
 & GPC & $-0.0046$ & $-0.0090$ & $[-0.0190, +0.0010]$ \\
\addlinespace
\multirow{2}{*}{ToN-IoT (constructed)} & SVC & $-0.0016$ & $-0.0011$ & $[-0.0049, +0.0028]$ \\
 & GPC & $+0.0313$ & $+0.0109$ & $[+0.0017, +0.0202]$ \\
\midrule
\multicolumn{2}{@{}l}{Source-dataset-equal mean, SVC} & $-0.0041$ & $-0.0056$ & $[-0.0070, -0.0045]$ \\
\multicolumn{2}{@{}l}{Source-dataset-equal mean, GPC} & $+0.0059$ & $-0.0009$ & $[-0.0019, +0.0009]$ \\
\bottomrule
\end{tabular}
\end{table*}

Two secondary checks do not alter this fixed-case conclusion. Source-dataset-equal estimates remain close across blocked, uniform, and kernel-stratified equal-count sampling (SVC $-0.00576$ to $-0.00548$; GPC $-0.00253$ to $-0.00185$). Observational nearest-rank pairings at the prespecified 1.25 caliper retain 75.2\% of pairs and have negative median quantum-minus-classical differences in all eight groups; one-to-one and caliper sensitivities agree. Supplementary Tables~9--11 and Supplementary Figure~4 give the complete schemes, counts, and cautions; none identifies a causal family effect.

\subsection{The quantum pool mixes entangling and product kernels}
\label{sec:res_circuits}

Circuit decomposition changes the interpretation of the aggregate family. The two ZZFeatureMap variants contain all-to-all two-qubit Pauli interactions, whereas the Pauli-X/Z and ZFeatureMap variants contain only one-qubit Pauli strings. For the latter pair, the prepared state and its fidelity kernel factorize over coordinates (Methods), permitting direct classical evaluation as a product of one-qubit overlaps. Thus two of the four map families cannot supply evidence that entanglement or hard quantum kernel estimation is responsible for their behaviour.

Table~\ref{tab:circuits} reports logical, unrouted resource counts. At 12 qubits, one ZZFeatureMap state preparation requires 132 CX gates for one repetition and 264 for two; the corresponding compute--uncompute fidelity templates require 264 and 528 CX gates. Both product-map templates require zero CX gates at every studied dimension. Under P1$'$ selection, product maps supply 594/1,080 (55.0\%) of the selected quantum SVC configurations and 348/1,080 (32.2\%) of the GPC selections. The composition alone does not compare predictive quality; the matched stratum analysis below does so at 30 candidates per stratum.

At matched 30-candidate budgets, the source-dataset-equal SVC effect is $-0.00928$ for the entangling-ZZ stratum and $-0.00547$ for the product-map stratum; the constituent source-dataset means range from $-0.01158$ to $-0.00580$ and from $-0.01167$ to $+0.00021$, respectively. The entangling-ZZ point estimate is negative in every fixed scenario-group ($-0.01960$ to $-0.00132$); product-map estimates range from $-0.02245$ to $+0.00571$. The full 60-candidate pool remains at $-0.00566$. Thus the controlled result is not produced by pooling product maps with entangling circuits, and restricting the quantum family to ZZ interactions does not reveal a hidden positive family effect. Supplementary Table~15 reports every pointwise conditional interval; these strata are fixed-case sensitivities, not new population tests.

\begin{table*}[t]
\centering
\caption{Logical circuit audit for the four quantum feature maps. Depth and CX columns give the values at 4 and 12 qubits after Qiskit 2.3.1 decomposition to \texttt{rz}, \texttt{sx}, \texttt{x}, and \texttt{cx}, optimization level 0, without device routing. ``Fidelity CX'' counts the compute--uncompute template for one kernel entry. Complete values at all five dimensions and one-qubit gate counts are provided in Supplementary Table~20 and the machine-readable artifact. These logical counts are not hardware-execution estimates.}
\label{tab:circuits}
\scriptsize
\setlength{\tabcolsep}{4pt}
\resizebox{\textwidth}{!}{%
\begin{tabular}{@{}llllccc@{}}
\toprule
Map & Pauli strings & Stratum & Product factorization & map depth (4/12) & map CX (4/12) & fidelity CX (4/12) \\
\midrule
ZZ, 1 rep. & Z, ZZ & entangling ZZ & no  & 19/67  & 12/132 & 24/264 \\
ZZ, 2 reps. & Z, ZZ & entangling ZZ & no  & 35/107 & 24/264 & 48/528 \\
Pauli-X/Z, 1 rep. & X, Z & separable product & yes & 11/11 & 0/0 & 0/0 \\
Z, 2 reps. & Z & separable product & yes & 8/8 & 0/0 & 0/0 \\
\bottomrule
\end{tabular}
}
\end{table*}

\subsection{Evaluation choices interact within the controlled protocol}
\label{sec:res_v8_factorial}

\begin{figure*}[t]
\centering
\includegraphics[width=\textwidth]{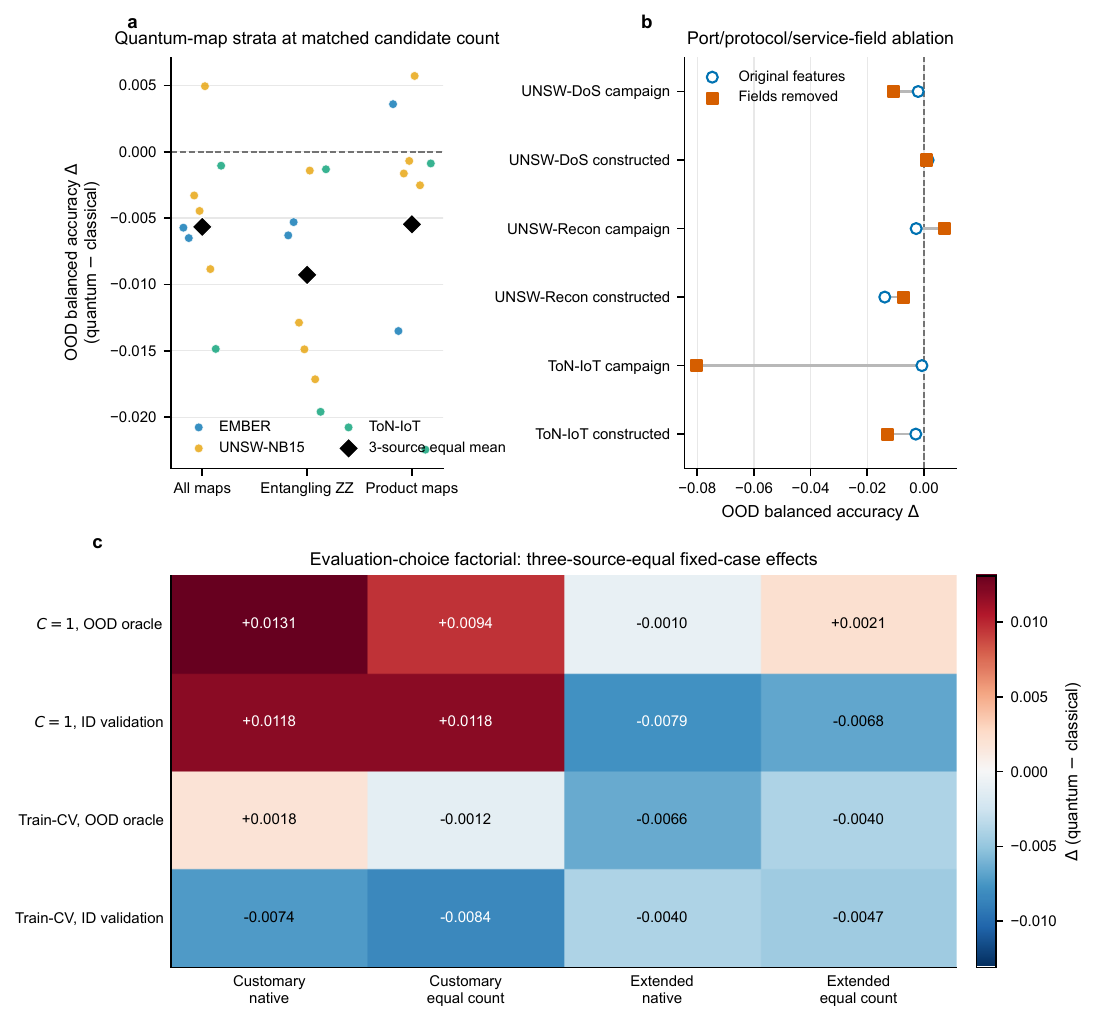}
\caption{Circuit-aware and evaluation-choice fixed-case sensitivities for SVC. \textbf{a}, Quantum-minus-classical OOD balanced-accuracy effects for all quantum maps ($B=60$), entangling-ZZ maps ($B=30$), and separable-product maps ($B=30$), each against a blocked equal-count extended-classical reference. Coloured points are the eight scenario-groups and black diamonds are three-source-dataset-equal means. \textbf{b}, Original and ablated controlled effects after removing the frozen port-, protocol-, and service-related fields in the six q1000 network groups. Connecting lines show paired group changes. \textbf{c}, Three-source-dataset-equal effects for the q1000 evaluation-choice $2\times2\times2\times2$ factorial. Positive values favour the fidelity-kernel family. All quantities are descriptive summaries conditional on the fixed datasets and finite candidate pools; panels do not show population uncertainty or causal attribution.}
\label{fig:v8_sensitivities}
\end{figure*}

The evaluation-choice factorial holds the data generation fixed while varying four evaluation choices. In the q1000 SVC stratum, the fixed-$C$, same-test-oracle, customary-reference, native-budget endpoint is $+0.0131$, whereas the train-CV, ID-validation, extended-reference, equal-count endpoint is $-0.0047$, a descriptive contraction of $0.0178$. The 16 three-source-dataset-equal endpoints range from $-0.0084$ to $+0.0131$ (Figure~\ref{fig:v8_sensitivities}; Supplementary Table~17).

The largest paired mean changes arise from train-CV regularization ($-0.0084$) and the extended reference ($-0.0079$), but a regularization-by-reference interaction of $+0.0139$ prevents additive or causal attribution (Supplementary Table~18). These controls therefore interact and must be reported together rather than attributed individually.

Removing the frozen port-, protocol-, and service-related fields also changes the relative comparison, confirming that common feature access alone does not make shortcut use family-invariant. Across the two network source datasets, the controlled effect moves from $-0.0031$ to $-0.0245$ (change $-0.0215$; source-dataset changes $-0.0447$ for ToN-IoT and $+0.0018$ for UNSW-NB15). The six group changes range from $-0.0794$ to $+0.0099$, and every pointwise change interval includes zero (Supplementary Table~19). The large negative change is concentrated in the ToN-IoT campaign shift, where removing direct traffic descriptors reduces both families' absolute accuracy but the selected fidelity kernel more strongly. This sensitivity neither restores a positive aggregate quantum effect nor establishes that all shortcuts have been removed.

\subsection{Protocol sensitivity is prospectively corroborated on external domain shifts}
\label{sec:res_external}

\begin{figure*}[t]
\centering
\includegraphics[width=\textwidth]{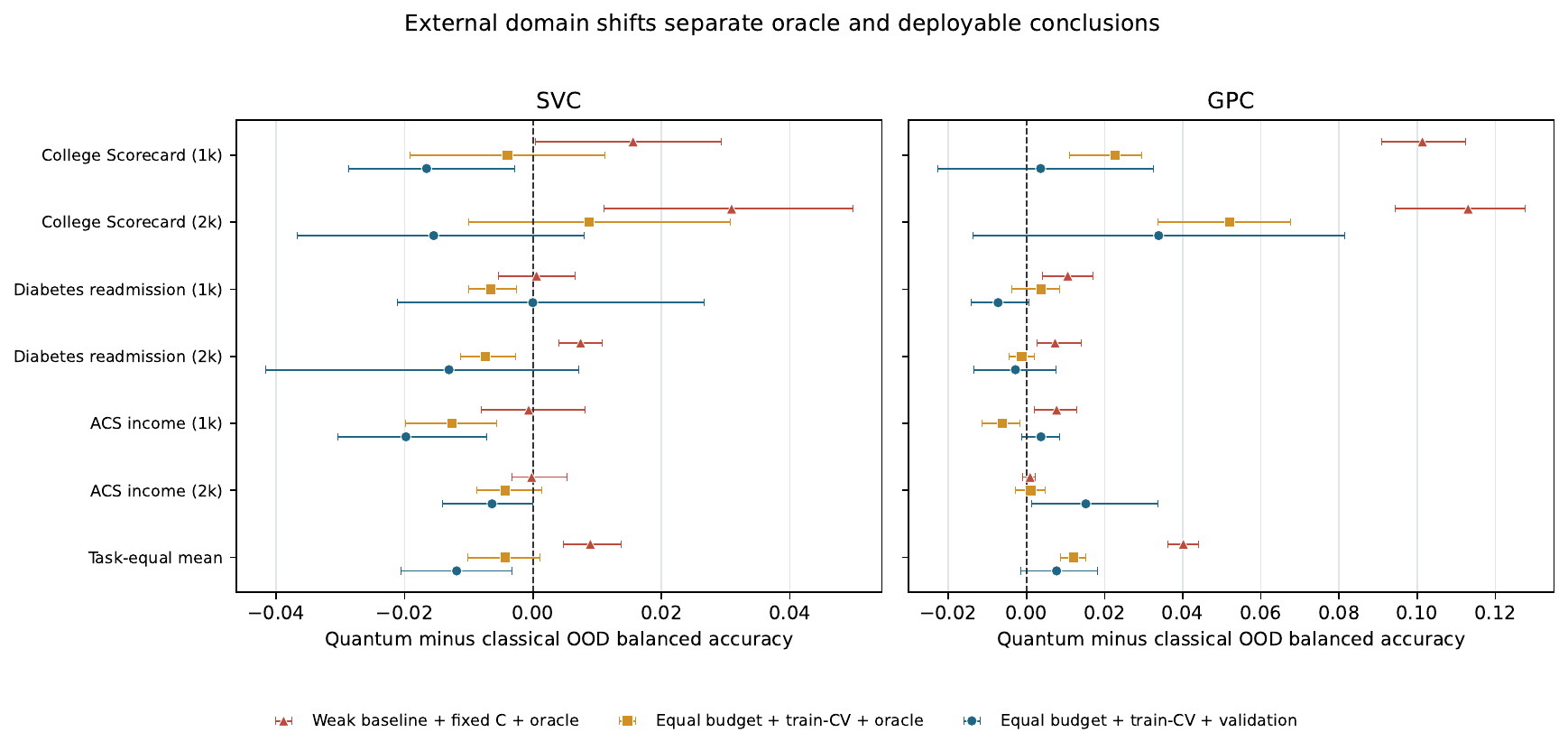}
\caption{External domain shifts separate optimistic and deployable conclusions. Effects are quantum minus classical OOD balanced accuracy for the weak-reference fixed-regularization same-test oracle, the equal-budget train-tuned same-test oracle, and the primary equal-budget train-tuned validation selector. Rows show both nested sizes for all three fixed tasks plus the task-equal mean. Error bars are 95\% conditional seed-cluster intervals over five deterministic subsampling seeds per task ($n=5$), preserving the nested-size relation. They quantify pipeline variation conditional on the fixed tasks and design, not uncertainty over a population of tasks.}
\label{fig:external}
\end{figure*}

We next applied the prospectively frozen protocol to College Scorecard, diabetes readmission, and ACS income, preserving the published TableShift domain splits and withholding all OOD labels from the controlled selector. The complete external grid contains 30 task--size--seed units and 37,800 split-level results; every prespecified unit passed the automated author-provided audit. Figure~\ref{fig:external} shows all task--size effects and conditional seed-cluster intervals.

The primary SVC result satisfies the machine-readable outcome category frozen as \texttt{replicated\_protocol\_sensitivity}; narratively, protocol sensitivity is prospectively corroborated across three fixed external tasks. The task-equal controlled effect is $-0.0119$ (95\% conditional seed-cluster interval $[-0.0205,-0.0033]$, $n=5$ deterministic subsampling seeds per task), whereas the weak-reference, fixed-$C$, same-test-oracle effect is $+0.0089$ ($[+0.0047,+0.0137]$). Their contraction is $+0.0208$ ($[+0.0113,+0.0307]$), with a leave-one-task-out range of $[+0.0116,+0.0260]$. Size-equal controlled point estimates are $-0.0160$ for College Scorecard, $-0.0066$ for diabetes readmission, and $-0.0131$ for ACS income; their pointwise intervals and both nested sizes are shown in Figure~\ref{fig:external}.

The secondary GPC shows a larger contraction of $+0.0324$ ($[+0.0223,+0.0427]$), but its task-equal controlled effect is $+0.0077$ with an interval spanning zero ($[-0.0015,+0.0183]$). College Scorecard leans quantum under this classifier, diabetes leans classical, and ACS is slightly positive. This classifier dependence limits the conclusion appropriately: the external study corroborates the sensitivity of the family verdict to evaluation design, not universal classical dominance.

\subsection{Geometry is associated with robustness only partially}
\label{sec:res_mechanism}

Effective rank and OOD kernel-target alignment are associated with OOD accuracy, but the association is regime-dependent. Across scenario-groups, median Spearman correlations span $-0.15$ to $+0.80$ for effective rank and $+0.22$ to $+0.64$ for OOD alignment; a leave-one-source-dataset-out ranking model reaches only $0.22$--$0.40$. Because OOD alignment reuses the labels on which accuracy is measured, it is a post hoc diagnostic. Supplementary Figure~5 and Supplementary Table~11 report the full analysis; neither descriptor is interpreted as a causal mediator or deployable selector.

\subsection{Finite shots can create distinctness without useful advantage}
\label{sec:res_shots}

All primary quantum kernels use exact statevectors. To test whether finite estimation can create apparent non-emulability, we apply the same zero-label certificate to 30 repeated binomial fidelity estimates at 128, 512, 2,048, and 8,192 shots for one fixed SVC quantum model per case. Each replicate reselects $C$ from training data under unprojected, independently projected, and coherent Nystr{\"o}m-PSD conditions (Methods).

\begin{figure*}[t]
\centering
\includegraphics[width=0.86\textwidth]{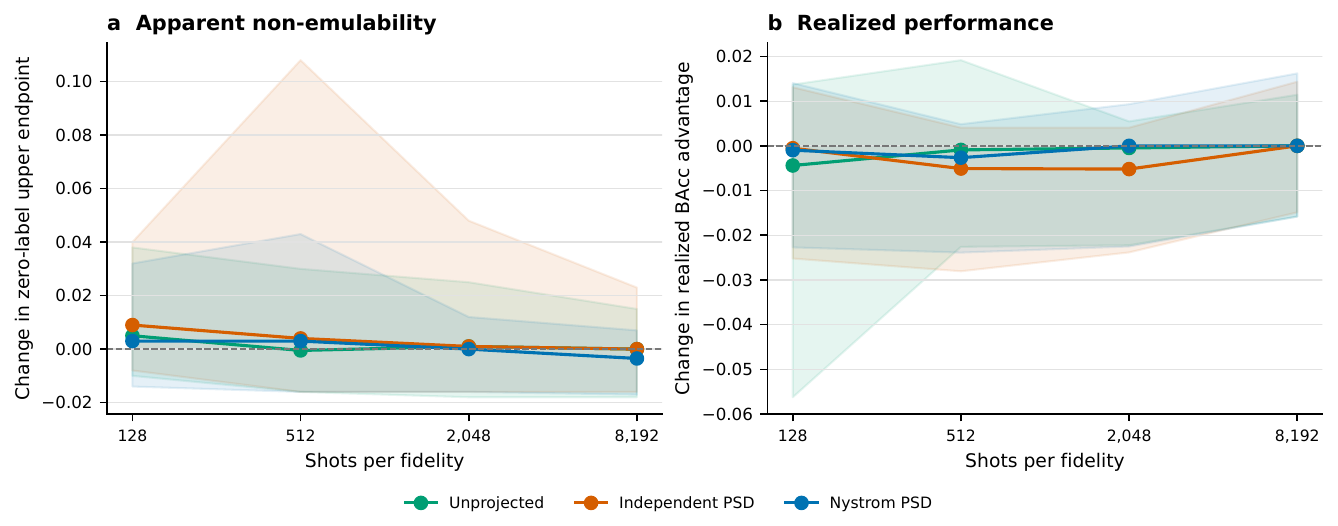}
\caption{Finite-shot perturbations can increase apparent non-emulability without increasing performance. Lines are medians across the eight fixed case-level Monte Carlo medians and ribbons span their minima and maxima. \textbf{a}, Change from the exact-statevector zero-label accuracy upper endpoint against the full classical family. \textbf{b}, Change in realized OOD balanced-accuracy advantage. Each case has 30 measurement replicates at each shot level and PSD condition. Ribbons are fixed-case ranges, not confidence intervals.}
\label{fig:v9_shots}
\end{figure*}

At 128 shots the perturbed model disagrees with its exact-statevector counterpart on a median $7.2$--$7.4\%$ of target decisions, depending on PSD handling. Yet the across-case median increase in the full-family zero-label endpoint is only $+0.003$ for Nystr{\"o}m, $+0.005$ unprojected, and $+0.009$ for independent projection; individual cases range from decreases of $0.014$ to increases of $0.040$. The corresponding median realized balanced-accuracy changes are $-0.0010$, $-0.0044$, and $-0.0005$, respectively (Figure~\ref{fig:v9_shots}). As shots increase, median certificate changes generally approach zero, while the case ranges remain heterogeneous.

Finite-shot noise can therefore move a quantum predictor away from every exact classical witness without making it better. This is a specifically important insufficiency result for the certificate: predictive non-emulability is necessary for a material advantage against the reference family, but it is not sufficient, and measurement noise can manufacture distinctness. Effective-rank inflation and the original performance sensitivities remain in the Supplementary Information. One case--replicate requires 1,624,250 distinct fidelity estimates; the full four-shot-level sensitivity projects 4.24 trillion circuit shots before routing and device overhead, halved if the four selected product-map cases are evaluated by direct factorization instead.

\section{Discussion}
\label{sec:discussion}

The central result is an exact answer to a deliberately narrower question than quantum advantage in the complexity-theoretic sense: given fixed target inputs, fixed predictions, a bounded loss, and a prespecified classical-kernel reference family, how much predictive advantage remains compatible with the labels not yet inspected? The finite max--min gives the interval hull of the identified set after any partial audit without a sampling, calibration, prevalence, or shift assumption. No interval based only on those observables can be uniformly smaller. For zero--one accuracy, the disagreement matrix collapses that optimization to a closed form. This turns ``a classical model looks similar'' into a sharp and minimal finite-batch falsification statement with an explicit materiality threshold.

This scope also separates the result from certifiable-surrogate generalization bounds, which control a target model's population risk through a surrogate with a generalization guarantee and unlabeled disagreement \cite{bazinet2026bound}. Our models remain fixed, the estimand is advantage against the best member of a declared family, and every unaudited label completion remains admissible. The acquisition rule is therefore secondary: it chooses which ambiguity to remove, while the certificate is pathwise exact after any realized audit subset. In QML, this target-decision object complements Huang's train-sample geometric sieve rather than replacing it \cite{huang2021powerofdata}.

The empirical frontier is much sharper than the benchmark effect alone. Against 115 classical kernels, at most 33 of 500 labels under the fixed adaptive rule are required to rule out an accuracy advantage above one point in every retrospective SVC and GPC case. Strong prediction-aware controls are at least as efficient on average, with median 12.5 labels versus 13, whereas an exact retrospective oracle needs median 6.5. This negative algorithmic result matters: most of the gain comes from adjudicating actionable disagreements, not from a privileged acquisition heuristic. The same certificate contraction holds for the selected entangling-ZZ subset. In the prospectively specified replication, all four eligible task--classifier medians require at most three labels and the overall seed-level median is zero; the controls again show no acquisition-heuristic advantage. It does not establish universal classical emulation: it identifies the quantum model's decisions relative to the prespecified classical-kernel reference family on these particular target covariates. Its value is that every remaining ambiguity is visible in the certificate rather than hidden inside an average performance comparison.

The three-gate interpretation prevents stronger conclusions from being conflated. Protocol validity asks whether the reported comparison could be deployed; predictive indispensability asks whether a witness in the prespecified classical-kernel reference family reproduces the fixed target decisions; quantum relevance asks whether any surviving behaviour depends on a non-factorizable circuit and remains meaningful after accounting for estimation cost. Passing all three gates creates a candidate for deeper advantage analysis, not a proof of speedup. Here Gate~1 reverses the optimistic sign, Gate~2 rules out a material batch advantage with limited auditing, and prospective corroboration in two technically eligible tasks meets the predefined criteria. Gate~3 finds no rescue from the retrospective entangling stratum while finite-shot noise sometimes increases distinctness without performance. Because all prospective winners are product maps, the prospective experiment corroborates Gate~2 but adds no independent evidence about Gate~3.

From the perspective of distribution-shift research, this reinforces that robustness conclusions are conditional on the protocol \cite{quinonero-candela2008datasetshift,gulrajani2021domainbed,koh2021wilds,yao2022wildtime}. Within the controlled protocol, moving from the fixed-$C$, same-test-oracle, customary-reference, native-budget corner to the fully controlled corner changes the q1000 estimate from $+0.0131$ to $-0.0047$. Train-CV regularization and the extended reference have the largest mean paired changes, but their sizeable interaction precludes an additive or causal reading. The cross-protocol display cannot isolate regularization either, so the larger $+0.037$ endpoint obtained under oracle selection with fixed regularization is interpretable only as the outcome of a complete evaluation protocol, not of any single choice.

The external analysis provides prospective corroboration beyond the constructed security shifts. Under a protocol frozen before result inspection, three education, health, and socioeconomic tasks corroborate a positive protocol contraction and a negative controlled SVC effect. The result is stable to omitting any one task, yet heterogeneous by task and classifier: the GPC endpoint retains a small positive task-equal lean whose interval crosses zero. The strongest defensible conclusion is therefore about evaluation sensitivity, not the universal ordering of kernel families.

Length scale is decisive for both classical and quantum kernel behaviour \cite{shaydulin2022bandwidth,canatar2023bandwidth}. With symmetric scale freedom and tuned regularization, the families occupy overlapping effective-rank and alignment regimes. These quantities are kernel descriptors, not unique signatures of quantum provenance; OOD alignment additionally reuses target labels and remains strictly post hoc.

The high-rank finding must also be reconciled with two spectral results that superficially point the other way. K\"ubler et al.\ argue that quantum kernels generalize in-distribution only when the relevant spectral mass is concentrated in few components \cite{kubler2021inductivebias}, and Thanasilp et al.\ show that fidelity kernels concentrate exponentially---toward effectively rank-one, uninformative Gram matrices---as qubit counts grow \cite{thanasilp2024concentration}. There is no contradiction, but there is a genuine tension that our results locate precisely. Both results concern the asymptotic or i.i.d.\ regime; our effective ranks (roughly $10$--$150$ on training sets of $10^3$--$4\times10^3$ samples at $4$--$12$ qubits) sit far below the concentration regime. The repeated finite-shot sensitivity adds an empirical caveat these theoretical results anticipate: effective rank from sampled fidelities is inflated by a median factor of $1.93$ at 128 shots and $1.13$ at 8,192 shots (Section~\ref{sec:res_shots}; Supplementary Information, repeated finite-shot sensitivity). Exact effective rank is therefore a statevector idealization even at these modest qubit counts.

From a practical perspective, no consistent fidelity-kernel advantage is observed over the tuned classical family, while overlap estimation introduces circuit and shot costs absent from direct classical-kernel evaluation. A shortened-length-scale Laplacian is therefore an important inexpensive reference, not a universally superior choice. More generally, comparisons should keep representation and learner fixed, tune all regularization on training data, match search budgets, select without target labels, include a non-constructed shift where possible, and report correlated benchmarks as fixed cases; a recent power-system attack-detection benchmark independently reports that evaluation-protocol and tuning choices each moved or reversed a quantum-versus-classical conclusion at fixed models \cite{islam2026benchmarking}. Geometry descriptors remain useful diagnostics, but not deployable selectors unless validated without target-label access.

Several limitations define the scope of our conclusions.

Our shift operationalization covers three reproducible mechanisms---within-class sparsity-tail extremes, train-geometry-tail extremes, and a held-out attack-campaign shift on network traffic (an unseen attack campaign plus a capture-partition change, not a timestamped temporal split)---but not the full space of real-world drift processes. Explicit temporal splits on timestamped malware, family-based drift, and adversarially induced shift remain outside the study's scope, and our conclusions should be re-tested under them rather than assumed.

Kernel and encoding coverage also remains finite. The classical family spans seven kernels plus a bandwidth sweep, and both families receive symmetric tuning freedom, but neither candidate set is exhaustive: the quantum side is restricted to four commonly used fidelity feature maps at low qubit counts, and trained or task-adapted encodings, projected kernels, and deeper maps are not evaluated. The observed association suggests that future extensions should report their Gram-matrix geometry as well as provenance, but it does not predict their outcome. At substantially larger qubit counts, exponential kernel concentration \cite{thanasilp2024concentration} creates an additional regime not represented here, so our conclusions should not be extrapolated beyond the low-qubit study.

The certificate is relative to a finite, prespecified classical family and a fixed observed target batch. It does not rule out a quantum advantage against a different classical family, establish a surrogate outside the observed covariates, or convert candidate count into computational complexity. Likewise, the batch interval is not a confidence interval for a future deployment population. The optional population result requires i.i.d.\ target sampling and models fixed before observing that sample, assumptions that do not describe the constructed tail batches.

The partial-label result is mathematically valid for any audit subset, but the security acquisition curve is retrospective. In the separate prospective Gate-2 experiment against the prespecified classical-kernel reference family, tasks, prediction locks, policies, thresholds, and numerical criteria were fixed and hashed before label opening. The two technically eligible tasks met those predefined criteria. Its scope remains narrow: NHANES failed a prespecified feature gate, the seeds are deterministic subsamples rather than independent task replicates, and every prospective quantum winner is a product map. The result corroborates the certificate and actionable-disagreement principle on these fixed tasks, not a population claim, entangling-kernel replication, or superiority of bottleneck coverage. Balanced-accuracy certification additionally requires a prespecified prevalence or prevalence range; using the retrospectively observed prevalence is only a post-unlock sensitivity.

The quantum kernels are simulated rather than executed on hardware. The study therefore characterizes fidelity-kernel geometry, not noisy-device behaviour. Repeated binomial sampling quantifies one component of measurement uncertainty, but hardware noise, transpilation, readout, mitigation, and circuit-dependent correlations remain unmodelled \cite{wang2021nisq}. Finite-shot cost is not merely an implementation detail: Gentinetta et al. place kernel-estimation uncertainty inside the end-to-end complexity of quantum support-vector machines \cite{gentinetta2024complexity}, while exponential concentration can make informative spectral structure expensive to resolve at larger scales \cite{thanasilp2024concentration}. Adaptive allocation of a limited measurement budget across kernel entries has been reported to improve classifier fidelity and measurement efficiency relative to uniform allocation \cite{miroszewski2026adaptive}; our repeated-shot sensitivity uses uniform allocation and does not optimize it. Recent QML evidence is not uniformly negative. Agliardi et al. report a favorable 10-qubit hardware instance on real data (87\% accuracy versus 80\% for their classical benchmark) and comparable utility-scale results on 156-feature synthetic data (80\% versus 83\%) \cite{agliardi2026covariant}. Those study-specific, single-instance comparisons motivate continued hardware evaluation but do not alter our fixed-batch result or validate advantage under our shift protocol. We make no claim about hardware feasibility, resource advantage, or quantum speedup---the rigorously established quantum-classical learning separation concerns a specifically constructed problem, not natural data \cite{liu2021rigorous}. At the dimensions studied, the fidelity kernels are also classically simulable \cite{schreiber2023surrogates,jerbi2024shadows}, and the ZZ map admits a closed-form classical surrogate in the small-bandwidth regime \cite{tekeli2026zzsurrogate}; this is precisely what makes them available as practical kernel constructions today.

Dataset scope and feature caveats provide a final boundary. Dynamic malware analysis and encrypted traffic remain untested. The network exports retain port-derived counters and, for ToN-IoT, direct port, protocol, and service indicators that can support shortcuts in network intrusion detection system (NIDS) benchmarks \cite{engelen2021troubleshooting}. Sharing features controls the inputs but does not imply that distinct kernels exploit them equally. In the frozen q1000 ablation, removing these fields shifts the two-source-dataset-equal effect from $-0.0031$ to $-0.0245$, but the change is highly heterogeneous and concentrated in ToN-IoT. This is evidence that feature access can affect the relative comparison, not proof that every shortcut has been removed or that the ablated tasks define a preferred deployment benchmark.

We do not claim universal classical superiority, predictive equivalence, complexity-theoretic advantage, hardware relevance, or a refutation of quantum kernels. The proposed audit instead makes a candidate advantage progressively falsifiable: first by protocol controls, then by a prespecified classical witness family on target covariates, and finally by limited adjudication of the disagreements that keep a material advantage possible. Future work should broaden the prospective replication to more tasks and entangling winners, and pair candidate breadth with measured computational resources. This interpretation is consistent with calls for controlled QML benchmarking \cite{bowles2024betterthanclassical,schnabel2025scrutiny,kakavand2026benchmarking}.

\section{Methods}
\label{sec:methods}

\subsection{Problem setting and estimands}

We study binary security classification---malware versus benign software and attack versus benign traffic---under distribution shift. Let $D_{\mathrm{ID}}$ denote the in-distribution process and $D_{\mathrm{OOD}}$ a constructed stress-test or held-out-campaign process. Each run contains a training set $T$, an ID holdout that is divided into validation $V_{\mathrm{ID}}$ and test $S_{\mathrm{ID}}$, and an OOD test set $S_{\mathrm{OOD}}$. Preprocessing, samples, classifier implementation, tuning rule, and family-selection rule are shared; only the kernel changes. Comparisons are always within a classifier.

The primary endpoint is OOD balanced accuracy after P1$'$ deployment selection. A family effect is
\[
\Delta_{\mathrm{OOD}}=
\mathrm{BAcc}^{(Q)}_{\mathrm{OOD}}-
\mathrm{BAcc}^{(C)}_{\mathrm{OOD}},
\]
so positive values favour the quantum family. ID balanced accuracy and ID-to-OOD degradation are secondary. P2 selects from OOD information in other realizations, and P3 selects and evaluates on the same OOD test labels; both are explicitly non-deployable descriptive oracles.

\subsection{Sharp finite-target identification}
\label{sec:methods_identification}

For Theorem~\ref{thm:bounded_loss}, fix the prediction-dependent loss tables before the audit and write $a_{ij}(y)=\ell_{ij}^C(y)-\ell_i^Q(y)$. For any completion of the labels outside $L$,
\[
\Delta_\ell(y)=\frac1n\min_j\left\{S_j(L)+
\sum_{i\notin L}a_{ij}(y_i)\right\}.
\]
For the lower endpoint, minima commute and the label choices separate by item:
\[
\min_{y_{L^c}}\min_j\left\{S_j+\sum_{i\notin L}a_{ij}(y_i)\right\}
=\min_j\left\{S_j+\sum_{i\notin L}\min_y a_{ij}(y)\right\}.
\]
For the upper endpoint introduce one-hot variables $z_{iy}\in\{0,1\}$ with $\sum_{y\in\mathcal Y}z_{iy}=1$ and maximize $t$ subject to
\[
nt\leq S_j(L)+\sum_{i\notin L}\sum_{y\in\mathcal Y}
a_{ij}(y)z_{iy}\qquad\text{for every }j.
\]
This mixed-integer linear program enumerates the finite max--min implicitly. Finiteness of $\mathcal Y^{|L^c|}$ guarantees that both extrema are attained. Any interval based on the same loss tables and audited labels that excluded either endpoint would fail on the attaining completion, proving sharpness and minimality. Revealing a label restricts the completion set, which proves monotone contraction. The implementation is tested against exhaustive enumeration for binary Brier loss, arbitrary multiclass bounded-loss tables, arbitrary partial labels, and the zero--one specialization.

The empirical accuracy certificate changes the estimand from a selected family mean to a fixed-model comparison. Let $q_i=h_Q(x_i)$ and $c_{ij}=h_{C_j}(x_i)$ be hard predictions on a target batch of size $n$. The quantum model is the P1$'$ winner fixed using ID validation; the classical candidates, their train-only hyperparameters, and their target predictions are also fixed before any target-label audit. Every displayed certificate uses the same inherited canonical realization per security group: $n_{\rm train}=1000$, $n_{\rm ID}=500$, $n_{\rm OOD}=500$, master seed 42, q-split seed 42, and model seed 42. The exact scenario run prefixes and selected SVC/GPC configurations appear in Supplementary Table~3. For a candidate labelling $y$,
\[
g_j(y)=\operatorname{Acc}(q;y)-\operatorname{Acc}(c_j;y),
\qquad
\Delta_B(y)=\min_j g_j(y).
\]
Only disagreements contribute to $g_j$. If $d_j=n^{-1}\sum_i\mathbf 1\{q_i\neq c_{ij}\}$, then $g_j(y)\in[-d_j,d_j]$. The labelling $y_i=q_i$ attains $g_j=d_j$ for every $j$ simultaneously, so $\max_y\Delta_B(y)=\min_jd_j$. If $k$ maximizes $d_j$, the labelling $y_i=c_{ik}$ attains $\Delta_B(y)\leq g_k(y)=-d_k$; no $g_j$ can be below $-d_j\geq-d_k$, so $\min_y\Delta_B(y)=-\max_jd_j$. This proves Theorem~\ref{thm:zero_label} and its sharpness.

For an arbitrary audited subset $L$, define the observed signed contribution $S_j(L)$ and remaining disagreement count $R_j(L)$ as in Theorem~\ref{thm:partial_label}. Every completion satisfies
\[
\frac{S_j-R_j}{n}\leq g_j(y)\leq\frac{S_j+R_j}{n}.
\]
Assigning every unaudited label to $q_i$ attains all witness upper endpoints simultaneously. Assigning them to the witness that minimizes $S_j-R_j$ attains the lower endpoint of $\Delta_B$. This proves Theorem~\ref{thm:partial_label}. In binary classification, if $D_j$ is the original disagreement count and the audit has revealed $a_j$ disagreements favouring $q$ and $b_j$ favouring $c_j$, the interval simplifies to
\[
\left[
\min_j\frac{-D_j+2a_j}{n},\
\min_j\frac{D_j-2b_j}{n}
\right].
\]
Thus the upper endpoint cannot increase under arbitrary adaptive querying. A classical-favouring label on a binary disagreement reduces that witness endpoint by exactly $2/n$; a quantum-favouring label leaves its upper endpoint unchanged. These statements are deterministic for the observed batch and require neither i.i.d.\ sampling nor a model for the unaudited labels.

For completeness, an optional population form is available when the target inputs are i.i.d.\ and every model is fixed before observing them. A union bound and Hoeffding's inequality give, with probability at least $1-\delta$,
\[
\sup_y\Delta_B^{P_T}(y)
\leq \min_{j\leq M}\widehat d_j+
\sqrt{\frac{\log(2M/\delta)}{2n}}.
\]
We do not apply this result to the constructed tail batches, for which the finite-batch certificate is the appropriate estimand.

\subsection{Reference tiers and active target-label audit}
\label{sec:methods_frontier}

The partial-label audits use two nested operational tiers. The customary tier contains the 30 linear and RBF candidates; the full tier contains all 115 prespecified classical candidates. Separately, the frozen zero-label breadth sensitivity evaluates 30, 60, and 115 candidates using 5,000 nested whole-block orderings rooted at \texttt{ksf-v9-frontier-20260731}; a 60-candidate draw contains 12 of 23 five-dimension blocks and the full tier contains all blocks. Tier size measures reference breadth only. It is not a computational-resource claim because training time, memory, and model-construction cost are not equalized by candidate count.

At every audit step, we calculate the witness-specific numerator $S_j+R_j$. Among witnesses with unaudited disagreements, the adaptive coverage rule identifies those tying at the smallest reducible numerator and queries an unaudited input that covers the largest number of these bottleneck witnesses. Ties follow a fixed SHA-256 order. The rule uses target covariates, predictions, and previously revealed outcomes, but never an unrevealed label. It is evaluated retrospectively for thresholds $\tau\in\{0.005,0.010,0.020\}$.

We compare it with three stronger references. Dynamic random-active sampling selects by SHA-256 order among all unaudited points disagreeing with at least one current reducible bottleneck witness. Non-adaptive initial coverage ranks points once by how many zero-label bottleneck witnesses they cover, without label feedback. Both use 200 frozen SHA-256 tie orders. Finally, an undeployable label oracle gives the exact ex-post minimum number of queries over every possible audit subset. If $D_j$ is witness $j$'s total disagreement count, crossing threshold $\tau$ requires
\[
r_j(\tau)=\max\!\left\{0,\left\lceil\frac{D_j-n\tau}{2}\right\rceil\right\}
\]
revealed disagreements whose realized label favours witness $j$; the oracle minimum is the smallest feasible $r_j(\tau)$ over witnesses. This is exact because the family endpoint crosses when any one witness endpoint crosses. A prediction-independent complete ordering is retained only as a weak diagnostic. Comparator quantiles describe frozen algorithmic orders, not sampling uncertainty.

The zero-label SVC definition and go/no-go criterion were frozen before prediction outcomes were inspected. The arbitrary-partial-label acquisition extension was declared only after the target labels had been unlocked; the strong-comparator amendment was declared after the original acquisition curves were known. Both are exploratory on these eight cases. GPC repeats the mathematics and policies but is not an independent prospective validation. Target labels are used only to evaluate certificate contraction, the oracle, and the final realized effect; no kernel, classifier, regularization, threshold, or P1$'$ winner is refitted or reselected. Prediction archives and the separately stored label file are joined only after SHA-256, row-index, model, and target-length gates pass.

\subsection{Prospectively specified Gate-2 replication against the prespecified classical-kernel reference family}
\label{sec:methods_gate2_prospective}

The Gate-2 replication against the prespecified classical-kernel reference family was specified and hashed on 1 August 2026 before any of its three tasks was acquired, any model prediction was computed, or any target label was inspected. The source is the same TableShift implementation pinned at commit \texttt{fca9429814703a07e3902d005d46563a207b7f0a} \cite{gardner2023tableshift}. The task identifiers were BRFSS Diabetes (race shift), ACS Food Stamps (Census-division shift), and NHANES Lead (poverty shift). All three were attempted and none could be replaced based on availability or outcome.

For each task, seeds $\{42,123,999,7,2024\}$ define deterministic label-blind samples with $(n_{\rm train},n_{\rm val},n_{\rm ID},n_{\rm target})=(1000,250,250,500)$. The frozen train-only semantic encoding, scaling, TruncatedSVD dimensions $\{4,6,8,10,12\}$, and angular representation are reused verbatim. The 60 quantum candidates and 115 classical candidates, P1$'$ validation selection, candidate-specific train-CV for SVC, fixed Laplace GPC, and decision thresholds are unchanged. The customary 30-candidate tier is secondary; the full tier and $\tau=0.010$ define the prospective outcome. Secondary thresholds are 0.005 and 0.020.

The pipeline is fail-closed. A prediction-lock stage copies target labels to a separate sealed tree, removes them from the model state, and writes hard target predictions, row-order digests, selected validation-only quantum configurations, and SHA-256 hashes without calculating or printing target performance, prevalence, disagreement, or an acquisition curve. Candidate checkpoints contain only target predictions and validation-selection quantities. An aggregate manifest is accepted only when every available task--seed unit has all 175 kernel configurations for both classifiers or a permitted pre-model technical-failure record. The audit stage verifies that aggregate hash, records the one-time label opening and analysis-code hash, then joins labels by source-position digest. No model is refitted after the join.

For each classifier and the two classical tiers, the fixed policies are adaptive bottleneck coverage, 200 dynamic random-active SHA-256 orders, 200 non-adaptive initial-coverage orders, the exact ex-post label oracle, and a weak prediction-independent order. The hash root appends task, seed, model, tier, policy, and draw. The primary summaries are four or six task--classifier medians over five seeds, depending on technical availability. The predefined primary criterion requires every available task--classifier median to be at most 50 labels, the overall seed-level median to be at most 25, and no seed to exceed 100. With exactly two available tasks, all four cells are required and the result is explicitly technically limited. A secondary criterion requires all four medians to be at most 50 and an overall median at most 50; fewer than two available tasks is technically incomplete. The criteria were evaluated exactly once.

NHANES generated 11 non-constant encoded features in all five seeds, fewer than required for the frozen 12-dimensional candidate grid. Minimum-feature failure was an explicitly permitted technical-unavailability gate. No NHANES model candidate was executed, the task was not replaced, and dimensions were not changed. BRFSS Diabetes and ACS Food Stamps passed every gate, leaving ten task--seed units for the technically limited prospective outcome.

\subsection{Balanced-accuracy identification}
\label{sec:methods_bacc}

Balanced accuracy is not label-free unless target prevalence is known or restricted. At a fixed positive count $n_+$, we group inputs by their joint prediction signature $z=(q_i,c_{i1},\ldots,c_{iM})$. Let $n_z$ be the size of group $z$ and $k_z\in\{0,\ldots,n_z\}$ its unknown number of positive labels, with $\sum_zk_z=n_+$. For classifier $h$ with group-level prediction $h_z\in\{0,1\}$,
\[
\operatorname{BAcc}_h(k)=\frac12\left[
\frac{\sum_z k_z h_z}{n_+}+
\frac{\sum_z(n_z-k_z)(1-h_z)}{n-n_+}
\right].
\]
The sharp upper endpoint is the mixed-integer linear program that maximizes $t$ subject to
\[
t\leq\operatorname{BAcc}_Q(k)-\operatorname{BAcc}_{C_j}(k)
\quad\text{for every }j,
\]
together with the integer and prevalence constraints above. The lower endpoint is the minimum over witnesses of their corresponding linear minima. Relaxing $k_z$ to a continuous interval gives the sharp distributional or fractional-label LP. When prevalence is known only to lie in an interval, we maximize over its admissible integer counts. We report this as a sensitivity because the natural target prevalence is unavailable without labels; accuracy remains the main assumption-free certificate.

\subsection{Why train-only geometry cannot determine target prediction}
\label{sec:methods_transport}

The geometric difference and effective rank summarize training Gram structure, whereas deployment also depends on the target--train block. An explicit normalized PSD construction makes the distinction exact. Set $K_{TT}=I_n$. One valid joint Gram extension gives a unit-norm target feature orthogonal to all training features, hence $K_{UT}=0$. A second gives the target the same feature as the first training point, hence $K_{UT}=e_1^\top$. Both joint matrices are PSD, have unit diagonal, and share identical $K_{TT}$. For kernel ridge regularization $\lambda>0$, however, their target prediction operators are
\[
0
\quad\text{and}\quad
e_1^\top(I_n+n\lambda I_n)^{-1}
=\frac{e_1^\top}{1+n\lambda},
\]
whose spectral-norm difference is $1/(1+n\lambda)$. Train-only geometry can therefore screen possible separation but cannot characterize target transport. We use the hard-prediction certificate for SVC and GPC rather than treating this kernel-ridge construction as an endpoint result.

\subsection{Security datasets and shared split sizes}

The source datasets are EMBER \cite{anderson2018ember}, UNSW-NB15 \cite{moustafa2015unsw}, and ToN-IoT \cite{alsaedi2020toniot}. Four binary scenarios---EMBER malware, UNSW-NB15 DoS, UNSW-NB15 Reconnaissance, and ToN-IoT Scanning---produce two shift mechanisms each and hence eight scenario-groups. Classical and quantum kernels always receive identical row indices and class balance.

The 72 principal settings cross eight groups with master seeds $\{42,123,999\}$ and $(n_{\mathrm{train}},n_{\mathrm{ID}},n_{\mathrm{OOD}})$ equal to $(1000,500,500)$, $(2000,1000,1000)$, or $(4000,1800,1800)$. Each setting is instantiated with q-split seeds $\{42,123,999,7,2024\}$ and model seeds $\{42,123,999\}$, yielding fifteen pipeline realizations arranged in five q-split clusters. Indices are stored explicitly; automated gates verify split-role disjointness, required cardinalities, both classes, identical family inputs, and absence of label-derived features.

\subsection{EMBER shift construction}

EMBER uses the histogram and byte-entropy feature blocks and two label-conditional tail stress tests. They are not described as certified covariate shifts because conditioning the split on class does not establish invariance of $P(Y\mid X)$. For $m1$, the sparsity score is the number of non-zero entries in the selected feature block. Within each class, low- and high-score extremes supply the exact-size OOD set; training and ID samples are stratified draws from the remainder.

For $m2$, the final training set is fixed first. A MaxAbs scaler and TruncatedSVD representation are fitted on training data only, and a centroid is computed for each class. Every non-training sample is scored by Euclidean distance to its own-class centroid. Within-class low- and high-distance extremes form $S_{\mathrm{OOD}}$, and the ID holdout is sampled from the remaining pool. This construction probes unusual geometry relative to the fixed training set without using test information to fit the representation.

\subsection{Network-flow shift construction}

Numeric preparation removes identifiers and explicit label derivatives, leaving 39 UNSW-NB15 and 88 ToN-IoT features. The UNSW-NB15 export excludes direct categorical protocol, service, and state fields but retains three port-derived counters; ToN-IoT retains two port fields plus protocol and service indicators. A leakage audit records every exclusion and feature--label association; the largest observed absolute correlation is 0.62 and belongs to a retained traffic feature. Each scenario has reference and current pools and uses two mechanisms.

The \emph{m2-centroid} mechanism is the network analogue of EMBER $m2$: train-only MaxAbs+TruncatedSVD class centroids define within-class distance tails, which supply the OOD set. The \emph{attack-campaign} mechanism performs no geometric selection. Training and ID pools contain benign traffic and all non-target attack categories; OOD contains benign traffic and the held-out target campaign. UNSW-NB15 reference and OOD pools additionally use different official capture partitions. This is an unseen-campaign plus capture-partition shift, not a timestamped temporal split.

\subsection{Shared preprocessing and representations}

For every run and dimension $d\in\{4,6,8,10,12\}$, a MaxAbs scaler and TruncatedSVD are fitted on $T$; the reduced coordinates are standardized and mapped linearly with training-fitted parameters so that the \emph{training} coordinates lie in $[0,\pi]$. Held-out coordinates receive the same affine map and may extrapolate outside that interval under shift. The same embedded samples are supplied to both families, making the comparison a kernel swap rather than a representation swap. Dimension is a candidate axis. No validation, ID-test, or OOD-test row fits a transformation. The SVD random state equals the model seed.

\subsection{Classifiers and train-only regularization}

The primary learner is an SVC with a precomputed Gram matrix and balanced class weights \cite{cortes1995svm}. For every kernel geometry, $C\in\{0.01,0.1,1,10,100\}$ is selected by five-fold stratified cross-validation on the training Gram matrix; ties favour the smaller $C$. When $n_{\mathrm{train}}>2000$, this selection step uses a class-stratified random-state-20260717 subsample capped at 2,000 rows, while the final model is always refitted on the complete training block. Because $C$ is resolved before family selection, its grid points are not counted as separate family candidates.

The secondary learner is a binary Gaussian-process classifier with a logistic link and Laplace approximation \cite{rasmussen2006gpml}. It consumes the same precomputed kernel, uses diagonal jitter $10^{-8}$, at most 100 Newton iterations, and convergence tolerance $10^{-6}$. The logistic--probit approximation supplies predictive probabilities; no post-hoc calibration is applied.

\subsection{Classical kernel family}

The customary reference contains a cosine-normalized linear kernel and RBF kernel. The extended family adds cosine-normalized degree-2 and degree-3 polynomial kernels with $\gamma=1/d$ and intercept 1, the Laplacian kernel $\exp(-\|x-x'\|_1/\ell_1)$, and Mat\'ern-$3/2$ and Mat\'ern-$5/2$ kernels. The base $\ell_1$ and $\ell_2$ scales are median pairwise Manhattan and Euclidean distances, respectively, calculated from at most 2,000 training rows with the model seed; the RBF base scale is fitted from training variance. RBF, Laplacian, and both Mat\'ern kernels use factors $\{0.1,0.3,1,3,10\}$ around these training-only bases. Deduplication yields 23 kernel-shape/scale blocks crossed with five dimensions, or 115 extended-classical candidates in full-coverage groups.

\subsection{Quantum fidelity-kernel family}

The quantum family uses exact statevector fidelities
\[
K_Q(x,x')=\left|\langle\phi_Q(x)\mid\phi_Q(x')\rangle\right|^2
\]
\cite{havlicek2019quantumenhanced,schuld2019featurehilbert}. Qiskit 2.3.1 implements four map blocks \cite{javadiabhari2024qiskit}: fully connected ZZ maps with one and two repetitions; a one-repetition Pauli map with single-qubit strings \texttt{X} and \texttt{Z}; and a two-repetition Z map. Only the ZZ maps contain two-qubit interactions. The latter two prepare product states,
\[
U(x)=\bigotimes_{j=1}^{d}U_j(x_j),\qquad
K_Q(x,x')=\prod_{j=1}^{d}
\left|\langle\phi_j(x_j)\mid\phi_j(x'_j)\rangle\right|^2,
\]
so the constructor argument \texttt{entanglement="full"} stored for the Pauli-X/Z map has no effect without a multiqubit Pauli string. The product can be evaluated classically in $O(d)$ one-qubit-overlap calculations. We call these the entangling-ZZ and separable-product strata. Each map uses angle scales $\{0.5,1,2\}$ and all five dimensions, giving 60 candidates (30 per stratum) at 4--12 simulated qubits. Classical bandwidth and quantum angle-scale freedom both control locality \cite{shaydulin2022bandwidth,canatar2023bandwidth}; no circuit is trained.

\subsection{Circuit and sampling-resource audit}

Circuit counts use Qiskit 2.3.1 and the logical basis \{\texttt{rz}, \texttt{sx}, \texttt{x}, \texttt{cx}\} at transpiler optimization level 0, without backend layout or routing. We report each feature-map template and the compute--uncompute fidelity template formed by composing the map at $x$ with the inverse map at $x'$ after assigning distinct numeric inputs. Counts are compiler- and basis-conditional logical resources, not device schedules.

For the repeated q1000 finite-shot sensitivity, each case and measurement replicate estimates the off-diagonal training triangle, the combined 500-row ID-holdout-to-training rectangle (partitioned downstream into validation and test roles), the OOD-test-to-training rectangle, and the off-diagonal OOD square triangle; diagonals are fixed to one. Thus one kernel configuration requires
\[
\binom{1000}{2}+500(1000)+500(1000)+\binom{500}{2}
=1{,}624{,}250
\]
distinct fidelity estimates per shot level and replicate. Multiplying this count by shots gives the idealized circuit-shot total. The three PSD conditions reuse the same sampled entries and add only classical post-processing. Routing, calibration, mitigation, queueing, and device noise are excluded.

\subsection{Target-label-free family selection}

The original ID holdout is divided deterministically with class stratification into $V_{\mathrm{ID}}$ and $S_{\mathrm{ID}}$. Within each class, global row indices are ordered by SHA-256 of the salt \texttt{ksf-v4-idsplit::} and index, then assigned alternately to validation and test. The halves are disjoint and approximately class-balanced. P1$'$ deploys, within each family, the candidate with highest $V_{\mathrm{ID}}$ balanced accuracy and reports it once on $S_{\mathrm{ID}}$ and $S_{\mathrm{OOD}}$. Ties follow lexicographic candidate order. OOD labels never enter regularization, preprocessing, or P1$'$ selection.

\subsection{Candidate-budget matching}

Best-of-family selection rewards a larger pool. The confirmatory coverage audit found 115 extended-classical and 60 quantum candidates in every one of the eight scenario-groups and both classifiers. The primary comparison therefore draws 60 candidates from each family in all 16 group--classifier cells. For each cell, 5,000 extended-classical subsamples are generated with seed 20260715. The primary \texttt{kernel\_blocked} scheme samples 12 of the 23 complete kernel-shape/scale blocks, each crossed with all five dimensions, to mirror the 12 quantum feature-map/angle-scale blocks crossed with the same dimensions. This matches both candidate count and block structure, although it does not make the two search spaces identical. \texttt{uniform} samples individual candidates, while \texttt{kernel\_stratified} distributes the draw across kernel-shape/scale strata. Each subsample selects its P1$'$ winner; the classical endpoint is the resampling expectation. The full 115-candidate pool is secondary. Resampling percentiles describe sensitivity to the sampled candidate set, not population uncertainty.

\subsection{Entangling and separable quantum strata}

The circuit-stratum sensitivity reuses all controlled-protocol summaries and compares three SVC pools under train-CV and P1$'$: the full 60-candidate quantum family, the 30 entangling-ZZ candidates, and the 30 separable-product candidates. The 60-candidate pool is compared with 12 complete classical blocks and each 30-candidate stratum with six complete classical blocks, always crossed with all five dimensions. Each classical endpoint is the expectation over 5,000 \texttt{kernel\_blocked} draws using SHA-256-derived seeds rooted at 20260729. This tests dependence on map stratum, not computational advantage.

\subsection{Evaluation-choice factorial and shortcut sensitivity}

The evaluation-choice SVC factorial uses all 360 q1000 runs and crosses: same-test OOD versus disjoint ID-validation selection; fixed $C=1$ versus train-CV $C$; customary versus extended classical reference; and native versus equal-count budgets. Fixed-$C$ predictions are recomputed with the deterministic controlled-protocol preprocessing, kernel constructors, samples, and split roles; an integrity gate requires exact candidate--split key identity with every archived controlled-protocol SVC summary and the original ID-split hash salt. With the customary 30-candidate reference, equal count samples six of twelve quantum blocks; with the extended reference, it samples 12 of 23 classical blocks. The same 5,000 frozen block draws are paired across selection and regularization cells. Sixteen fixed-case endpoints and their paired axis contrasts are descriptive; they do not identify causal effects.

The q1000 shortcut sensitivity uses the same design on all 270 network runs after removing features before the train-only embedding. UNSW-NB15 excludes \texttt{ct\_src\_dport\_ltm}, \texttt{ct\_dst\_sport\_ltm}, and \texttt{is\_sm\_ips\_ports}; ToN-IoT excludes source/destination ports, all \texttt{proto\_*}, \texttt{service}, and \texttt{service\_*} fields. This removes 3 of 39 and 14 of 88 fields, respectively. The manuscript-facing contrast uses train-CV, P1$'$, and 60-candidate block matching with paired draws. It is a port/protocol/service-field ablation, not a guarantee that every shortcut has been removed.

\subsection{Security fixed-case aggregation and conditional intervals}

The effective replication axis within a scenario-group is q-split seed. Master seed is a design stratum rather than a replicate, some EMBER OOD pools recur across master seeds, and training sizes are nested. Audited OOD overlap is at most 1.2\% for EMBER and 7.8\% for network groups. Within each q-split cluster, model seeds are averaged within master seed, master seeds within size, and the three sizes with equal weight. The five resulting cluster means $T_j$ give the group effect $\bar T$ and 95\% conditional cluster-$t$ interval
\[
\bar T\ \pm\ t_{4,0.975}\frac{s_T}{\sqrt{5}}.
\]
This few-cluster $t$ construction follows the cluster-mean approach of Ibragimov and M{\"u}ller \cite{ibragimov2016fewclusters}. The interval ($n=5$ q-split clusters) is conditional on the fixed benchmark pools and experimental design; inclusion of zero is not evidence of equivalence. A 9,999-draw bias-corrected and accelerated (BCa) bootstrap over cluster means and leave-one-q-split calculations are secondary diagnostics, reported for all 16 primary cells in Supplementary Table~13. The five cluster values behind every interval are exposed in Supplementary Table~14.

Across groups, EMBER $m1/m2$ are averaged within source dataset, the four UNSW-NB15 groups within UNSW-NB15, and the two ToN-IoT groups within ToN-IoT. These three source-dataset means receive equal weight. Leave-one-source-dataset-out ranges omit one of the three at a time. The eight scenario-groups and three datasets were not sampled from defined populations; therefore no population $p$-value or dataset-population confidence interval is reported.

\subsection{Specification curve}

The S1--S10 display is a prespecified cross-protocol sensitivity map over complete endpoints from the earlier fixed-$C$ protocol and the controlled protocol. S1--S4 use the earlier fixed-$C$ protocol and S5--S10 the train-CV controlled protocol, so their boundary cannot isolate regularization. The rows remain in their prespecified order and are never sorted by effect. Individual-axis interpretation instead uses the evaluation-choice factorial above. All summaries average realizations within setting and group, then groups within the three source datasets, and finally source datasets with equal weight; no specification-level hypothesis test is performed.

\subsection{Prospectively frozen external corroboration}

Before acquiring TableShift data or inspecting any external result, we committed the tasks, sampling, preprocessing, candidate pools, estimands, uncertainty procedure, and outcome-category rule. The source implementation is pinned to commit \texttt{fca9429814703a07e3902d005d46563a207b7f0a} \cite{gardner2023tableshift}. Three public tasks span distinct domains and published shift variables: College Scorecard (institution type), diabetes readmission (admission source), and ACS income (geographic region). TableShift's \texttt{train}, \texttt{validation}, \texttt{id\_test}, and \texttt{ood\_test} roles are preserved; \texttt{ood\_validation} is never loaded. College Scorecard is read from its public CC0 archive because the pinned downloader invokes authenticated Kaggle access; the pinned parser, schema, and domain split are unchanged. Source-file SHA-256 hashes and the container digest are recorded.

Each task has nested strata $(n_{\mathrm{train}},n_{\mathrm{val}},n_{\mathrm{ID}},n_{\mathrm{OOD}})=(1000,250,250,500)$ and $(2000,500,500,1000)$. For seeds $\{42,123,999,7,2024\}$, source rows are ordered by SHA-256 of task, split, seed, and original position, giving deterministic label-blind samples. Gates verify source-role disjointness, size nesting, both classes, source-schema preservation, and at least 12 non-constant encoded features. All 30 task--size--seed units passed.

All transformations fit \texttt{train} only. Constant columns are removed; numeric variables receive median imputation; Boolean, categorical, and string variables receive most-frequent imputation and one-hot encoding with unknown categories ignored. MaxAbs scaling, TruncatedSVD at the five study dimensions, standardization, and a training-fitted affine map that places training coordinates in $[0,\pi]$ follow; held-out coordinates may extrapolate beyond that range. The SVD random state is 20260729. The security kernel and classifier pools are then reused.

For each task, size, seed, and classifier, the validation-selected 60-candidate quantum endpoint is compared with 5,000 \texttt{kernel\_blocked} 60-candidate subsamples of the 115-candidate classical pool using seed 20260729. The primary effect is quantum minus expected classical OOD balanced accuracy. The equal-budget OOD oracle selects on \texttt{ood\_test}; the weak-fixed oracle uses linear/RBF, fixes SVC at $C=1$, and also selects on the reported OOD labels. Protocol contraction is weak-fixed-oracle effect minus controlled P1$'$ effect.

Seeds are averaged within task and size, nested sizes receive equal weight within task, and the three fixed tasks receive equal weight. Conditional 95\% intervals use 9,999 resamples of the five seed clusters within task while preserving nested sizes. Leave-one-task-out ranges are descriptive. The intervals quantify deterministic-subsampling variability conditional on three fixed tasks; they do not represent a population of tasks.

\subsection{Performance and uncertainty metrics}

Balanced accuracy on $S_{\mathrm{OOD}}$ is primary. Accuracy, macro and positive-class F1, ROC AUC, and precision--recall AUC are recorded on every split. The GPC analysis additionally records log loss, Brier score, expected calibration error with 15 equal-width probability bins \cite{guo2017calibration}, and mean predictive entropy. No threshold or calibration parameter is fitted on OOD labels.

\subsection{Geometry descriptors and association analyses}

For positive-semidefinite $K$ with eigenvalues $\lambda_i$ and $p_i=\lambda_i/\sum_j\lambda_j$, effective rank is
\[
r_{\mathrm{eff}}=\exp\!\left(-\sum_i p_i\log p_i\right)
\]
\cite{roy2007effective}. For $y\in\{-1,+1\}^n$ and doubly centered $K_c$, centered kernel-target alignment is
\[
\mathrm{KTA}(K,y)=
\frac{\langle K_c,yy^\top\rangle_F}{n\|K_c\|_F}
\]
\cite{cristianini2001kta,cortes2012centeredalignment}. KTA is computed separately on training, ID, and OOD square blocks, alongside off-diagonal and cross-split similarities. OOD KTA reuses the same OOD labels as accuracy and is therefore a post hoc diagnostic, not a label-independent mechanism.

Within each run--dimension cell, Spearman correlations relate effective rank, OOD KTA, and KTA survival (OOD minus ID) to SVC OOD balanced accuracy for all, classical-only, and quantum-only candidates. Partial Spearman association residualizes ranked variables against family and log length/angle scale. For portability, a descriptive model bins log effective rank into ten training-dataset quantiles, learns mean OOD accuracy per bin from two source datasets, and ranks cells in the third; held-out Spearman correlation is reported once per source dataset.

Rank matching is observational. Within each run and dimension, every quantum candidate is first paired to the classical candidate with smallest absolute effective-rank difference; classical candidates may be reused. Rank ratio $\max(r_Q,r_C)/\min(r_Q,r_C)$ is filtered at prespecified calipers 1.10, 1.25, 1.50, and 2.00, plus unfiltered reporting. A sensitivity uses minimum-cost one-to-one assignment with absolute log-rank difference, leaving excess candidates unmatched. Counts, retained fractions, rank discrepancies, mean and median OOD differences, and the fraction favouring quantum are reported for every caliper. None of these analyses identifies a causal geometry or family effect.

\subsection{Repeated finite-shot perturbation}

The finite-shot analysis and its out-of-sample PSD extension were each specified before their outcomes were computed. They fix one $q1000$, master-seed-42, q-split-42, model-seed-42 run per security group and, within each, the exact-statevector quantum SVC candidate selected by P1$'$ on $V_{\mathrm{ID}}$. Lexicographic candidate order resolves ties. Input \texttt{summary\_v4.csv} SHA-256 hashes, chosen kernels, dimensions, and exact $C$ values are stored in the frozen specifications.

For shots $s\in\{128,512,2048,8192\}$ and measurement replicates $r=0,\ldots,29$, each exact fidelity $p$ is sampled independently as $\operatorname{Binomial}(s,p)/s$. A separate NumPy generator is used for train, ID-validation, ID-test, OOD-test, and OOD-square blocks. Its unsigned 64-bit seed is the first eight SHA-256 bytes of
\[
\texttt{ksf-v6-shots::<run>::<kernel>::<dim>::<shots>::<replicate>::<block>}.
\]
Python's process-randomized hash is not used.

Square train and OOD blocks are sampled on the upper triangle, mirrored, clipped to $[0,1]$, and assigned unit diagonal; rectangular blocks receive binomial sampling and clipping. Three prespecified conditions use the same sampled blocks. The unprojected condition makes no PSD correction. The independent-square heuristic clips negative eigenvalues of the train and OOD square matrices separately while leaving every rectangular evaluation--train block unchanged; it is an algorithmic intervention for a precomputed SVC, not a coherent correction of one out-of-sample kernel.

The coherent Nystr{\"o}m condition eigendecomposes the sampled training matrix as $K_T=U\operatorname{diag}(\lambda)U^\top$ and retains $\lambda_i>\tau$, where $\tau=\max\{10^{-12},\epsilon n\max(1,\max_i|\lambda_i|)\}$. For retained $U_+$ and $\lambda_+$,
\[
K_T^+=U_+\operatorname{diag}(\lambda_+)U_+^\top,\qquad
\Phi_E=K_{E,T}U_+\operatorname{diag}(\lambda_+)^{-1/2}.
\]
The corrected evaluation--train and evaluation-square blocks are $\Phi_E\Phi_T^\top$ and $\Phi_E\Phi_E^\top$, respectively, with $\Phi_T=U_+\operatorname{diag}(\lambda_+)^{1/2}$. Thus the OOD square used for KTA is derived from the sampled OOD--train block rather than independently projected. In all three conditions, $C$ is reselected by the same five-fold train-only procedure and the final SVC is refitted. Outputs include signed and absolute OOD change from the fixed exact endpoint, effective-rank ratio, OOD-KTA change, retained rank, eigenvalue tolerance, negative-eigenvalue fraction, and Frobenius changes for every block. Within each fixed Gram matrix, medians and central 2.5--97.5\% Monte Carlo intervals summarize 30 replicates. Across the eight fixed matrices, medians and ranges are descriptive.

For the finite-shot certificate extension we export the hard target prediction of every perturbed SVC and compare it with the unchanged exact-statevector classical witnesses. For each replicate and classical tier we record self-disagreement from the exact quantum predictor, the zero-label interval, and realized accuracy and balanced-accuracy advantages after the integrity-gated label join. This deliberately asks whether quantum measurement perturbation creates behaviour absent from the existing classical reference; it does not give the classical family an additional noise or retuning budget. The model contains no device or correlated-noise process.

\subsection{Statistics and reproducibility}

No null-hypothesis test is used for the eight security groups, three source datasets, or three external tasks. Every uncertainty object names its effective $n$, resampling unit, and conditioning set: security intervals use five q-split clusters; external intervals use five deterministic seed clusters per task; finite-shot intervals use 30 measurement replicates conditional on one fixed Gram matrix. The fifteen security pipeline realizations are never treated as independent. Security intervals use the $t_4$ reference and are pointwise, not simultaneous; the five underlying cluster means are published. Inclusion of zero is never interpreted as equivalence. No multiplicity adjustment was prespecified. Because multiple intervals and diagnostics are displayed, they are interpreted descriptively rather than as a familywise error-controlled decision procedure; the BCa analysis with five clusters is retained only as a fragile sensitivity.

Sample sizes were design- and compute-constrained and fixed before outcome analysis; no power calculation was used. No completed group, task, specification, budget scheme, caliper, or finite-shot replicate was removed based on its result. Randomized operations use recorded seeds; external sampling, ID division, and audit tie-breaking use deterministic SHA-256 ordering. The test suite validates source-dataset mapping, endpoint equality across analyses, candidate-budget coverage, matching uniqueness, seed stability, exact-reference hashes, prediction/label separation, theorem endpoints by exhaustive enumeration, MILP consistency, monotone partial-label contraction, and complete Monte Carlo cells.

\subsection{Software environment}

Experiments use Python 3.11/3.12, NumPy 2.3.5, pandas 2.2.3, scikit-learn 1.8.0, Matplotlib 3.11.0, and Qiskit 2.3.1. The Windows-specific Conda snapshot and cross-platform package specifications serve different archival purposes; neither is an exact machine image of every full-scale run.

\subsection{Use of generative AI}

Large language models were used beyond copy-editing, and no model is an author. OpenAI Codex was used during manuscript revision for language editing, code inspection, analysis implementation, and consistency checking. Claude Code (Anthropic), running the Claude Fable 5.1 model, was used in September 2026 for editorial restructuring of the Introduction, literature search and reference formatting, adaptation to the journal format, and consistency checking. No model determined authorship, study design, data handling, analysis decisions, or the acceptance of any scientific claim. The authors inspected every generated change, verified all reported analyses and citations against primary sources, and take full responsibility for the manuscript.

\backmatter

\section*{Declarations}

\bmhead{Ethics approval and consent to participate}
This work is a secondary computational analysis of public benchmark releases. The authors recruited no participants, administered no intervention, collected no new human data, and accessed no direct identifiers. The cybersecurity benchmarks contain malware or network-flow records rather than human-participant observations. For the TableShift tasks, ACS PUMS records are public-use files with Census Bureau disclosure protection \cite{uscensus2026pums}; BRFSS survey data and documentation are publicly downloadable from CDC \cite{cdc2022brfss}; the NHANES source protocols are documented by the NCHS Ethics Review Board \cite{cdc2026nhanesethics}; and the diabetes-readmission records are distributed through the public UCI repository \cite{uci2014diabetes}. College Scorecard supplies public institution-level records \cite{used2026scorecard}. Raw source records are not redistributed. This study falls outside the scope of institutional ethics review because it consists exclusively of secondary analyses of publicly available, deidentified datasets, without recruitment, interaction, intervention, access to direct identifiers, or any attempt at reidentification. Accordingly, ethics approval and additional informed consent were not required. No ethics approval or exemption is claimed.

\bmhead{Consent for publication}
Not applicable.

\bmhead{Availability of data and materials}
The derived datasets supporting the conclusions of this article are available in the Zenodo repository, \url{https://doi.org/10.5281/zenodo.21776862} (immutable v1.1.6 release). This study uses the public EMBER \cite{anderson2018ember}, UNSW-NB15 \cite{moustafa2015unsw}, ToN-IoT \cite{alsaedi2020toniot}, and TableShift \cite{gardner2023tableshift} benchmarks, which remain subject to their original access conditions; raw source records are not redistributed. The archived release contains derived split definitions, deterministic row-position hashes, shift constructions, experiment manifests, complete aggregated results, target-prediction matrices, physically separated audit labels, evidence frontiers, technical-failure records, and integrity manifests. The all-version concept DOI is \url{https://doi.org/10.5281/zenodo.19147649}; the exact v1.1.5 predecessor is \url{https://doi.org/10.5281/zenodo.21764577}, the exact v1.1.4 predecessor is \url{https://doi.org/10.5281/zenodo.21759058}, and the exact v0.8.0 predecessor is \url{https://doi.org/10.5281/zenodo.21717074} \cite{fernandezbarrios2026artifactv08}.

Code for constructing the splits, running the kernel experiments, reproducing every analysis, and regenerating every table and figure is available as follows. Project name: kernel\_shift\_framework. Project home page: \url{https://github.com/roberto-fernandez-barrios/kernel_shift_framework}. Archived version: \url{https://doi.org/10.5281/zenodo.21776862}. Operating system(s): platform independent; a Windows-specific Conda snapshot and platform-neutral package specifications document the dependencies, and neither is claimed as an exact machine image of every full-scale run. Programming language: Python. Other requirements: Python 3.11 or later with NumPy, pandas, scikit-learn, Matplotlib, and Qiskit at the versions listed in the archived lock files. License: BSD-3-Clause. Any restrictions to use by non-academics: none.

\bmhead{Competing interests}
All authors declare no financial or non-financial competing interests.

\bmhead{Funding}
This study received no funding.

\bmhead{Authors' contributions}
R.F.-B.: conceptualization, methodology, software, data curation, formal analysis, visualization, investigation, writing---original draft, and writing---review and editing. I.P.-L.: supervision, validation, and writing---review and editing. A.G.-S.: validation, project administration, and writing---review and editing. P.G.B.: supervision, resources, and writing---review and editing. All authors read and approved the final manuscript.

\bmhead{Acknowledgements}
Not applicable.

\bmhead{Supplementary information}
Additional file 1: Supplementary Information (PDF). Extended results, complete sensitivity tables, the prospective Gate-2 replication tables against the prespecified classical-kernel reference family, geometry and finite-shot diagnostics, and the machine-readable result map supporting the main article.


\bibliography{sn-bibliography}

\end{document}